 \documentclass[12pt]{article}
 \usepackage[dvips]{graphicx}
 \usepackage{epsfig}
 \usepackage{xspace}
 \usepackage{amssymb}
 \usepackage{cite}
 \usepackage{color}
 \usepackage{pennames}
 {\end{list}}
 \newcounter{enumct}
 \newenvironment{Enumerate}{\begin{list}{\arabic{enumct}.}%
 {\usecounter{enumct}\setlength{\topsep}{0.2mm}%
 \setlength{\partopsep}{0.2mm}\setlength{\itemsep}{0.2mm}%
 \setlength{\parsep}{0.2mm}}}{\end{list}}
 \newcommand{\captive}[1]{\rule{5mm}{0mm}%
 \begin{minipage}{150mm}\caption[small]{#1}\end{minipage}}
\newcommand{\WWbar  }{\ensuremath{\mathrm{W^+W^-}}\xspace}
\newcommand{\al     }{\ensuremath{\alpha_{s}}\xspace}
\newcommand{\bbbar  }{\ensuremath{\mathrm{b\bar{b}}}\xspace}
\newcommand{\ccbar  }{\ensuremath{\mathrm{c\bar{c}}}\xspace}
\newcommand{\dWGG   }{\ensuremath{\Delta W_\GG/W_\GG}\xspace}
\newcommand{\dWeG   }{\ensuremath{\Delta W_\eG/W_\eG}\xspace}
\newcommand{\eb     }{\ensuremath{E_{\mathrm{b}}}\xspace}
\newcommand{\ee     }{\ensuremath{{\mathrm e}{\mathrm e}}\xspace}
\newcommand{\eG     }{\ensuremath{{{\mathrm e}\gamma}}\xspace}
\newcommand{\epem   }{\ensuremath{\rm e^+e^-}\xspace}
\newcommand{\emga   }{\ensuremath{\Pem\,\Pgg}\xspace}
\newcommand{\etag   }{\ensuremath{E_{\mathrm{tag}}}\xspace}
\newcommand{\ft     }{\ensuremath{F_{2}^{\gamma}}\xspace}
\newcommand{\ftc    }{\ensuremath{F_{2,\mathrm{c}}^{\gamma}}\xspace}
\newcommand{\gev    }{\ensuremath{\mathrm{GeV}}\xspace}
\newcommand{\gevsq  }{\ensuremath{\mathrm{GeV^2}}\xspace}
\newcommand{\GG     }{\ensuremath{{\gamma\gamma}}\xspace}
\newcommand{\lame   }{\ensuremath{\lambda_{\rm e}}\xspace}
\newcommand{\lumi   }{\ensuremath{{{\cal L}_{int}}}\xspace}
\newcommand{\mc     }{\ensuremath{m_\mathrm{c}}\xspace}
\newcommand{\mumu   }{\ensuremath{\mu^+\mu^-}\xspace}
\newcommand{\Pc     }{\ensuremath{P_\mathrm{c}}\xspace}
\newcommand{\qqbar  }{\ensuremath{\mathrm{q\bar{q}}}\xspace}
\newcommand{\qsq    }{\ensuremath{Q^{2}}\xspace}
\newcommand{\stogg  }{\ensuremath{\sigma_\zg}\xspace}
\newcommand{\ttag   }{\ensuremath{\theta_{\mathrm{tag}}}\xspace}
\newcommand{\zg     }{\ensuremath{{\Pgg\,\Pgg}}\xspace}

\begin{document}
 \sloppy
 \pagestyle{empty}
%
%
 \begin{center}
 {\LARGE\bf Two Photon Physics}
 {\LARGE\bf at a Future Linear Collider\footnote{Invited talk
  at the Workshop on photon interactions and the photon structure,
  10-13 September 1998, Lund, Schweden, to appear in the Proceedings.}
  }\\[10mm]
 {\Large Richard Nisius}
 \\[3mm]
 {\it CERN,}\\[1mm]
 {\it CH-1211 Gen\`eve 23, Switzerland}\\[1mm]
 {\it E-mail: Richard.Nisius@cern.ch}\\[20mm]
%
%
 {\bf Abstract}\\[1mm]
 \begin{minipage}[t]{140mm}
 Some general considerations on a future linear collider and 
 selected topics of two photon physics measurements which can be
 performed at such a collider are presented.
 This review discusses 
 the total photon-photon cross section,
 jet cross sections, 
 structure functions, 
 charm production, 
 the BFKL Pomeron,
 $W$ pair production, 
 and Higgs production.
 \end{minipage}\\[5mm]
 \rule{160mm}{0.4mm}
 \end{center}
%
%
\section{Introduction}
\label{sec:intro}
 With the advent of a future linear collider two photon physics 
 will be important for several reasons.
 Firstly the high centre-of-mass energy in the order of $0.5-2$~TeV 
 enables to extend the two photon physics measurements performed at LEP.
 Examples of such measurements are
 the measurements of the photon structure functions which can be performed 
 at much higher momentum transfer squared of the virtual photon and
 the measurements of the total photon-photon cross section and of 
 jet cross sections in photon-photon scattering which
 can be extended to larger invariant masses and jet transverse momenta.
 Secondly two photon physics gets extended to new channels, especially 
 if a photon linear collider, PLC, can be 
 build~\cite{GIN-8301-GIN-8401-TEL-9001-TEL-9501}.
 In such a case the linear collider will be a $W$-factory with millions
 of polarised $W$ pairs being produced per year. 
 Even more important will be the fact that the Higgs boson can be produced 
 in the photon-photon fusion process $\GG\rightarrow H$. 
 The study of this process will provide very valuable information on
 how particle masses are generated.
 Thirdly two photon physics processes will play an even more important role 
 as background for searches for new physics than they already do in 
 searches performed at LEP.
 The material presented here is a personal collection of topics 
 which I find interesting and it is not a complete survey of all
 topics under study. The review mainly relies on the work
 done within the DESY ECFA study groups on the physics potential
 of a future linear collider.
%
%
\section{The Instruments}
\label{sec:instru}
\begin{figure}[htb]
\begin{center}
{\includegraphics[width=0.9\linewidth]{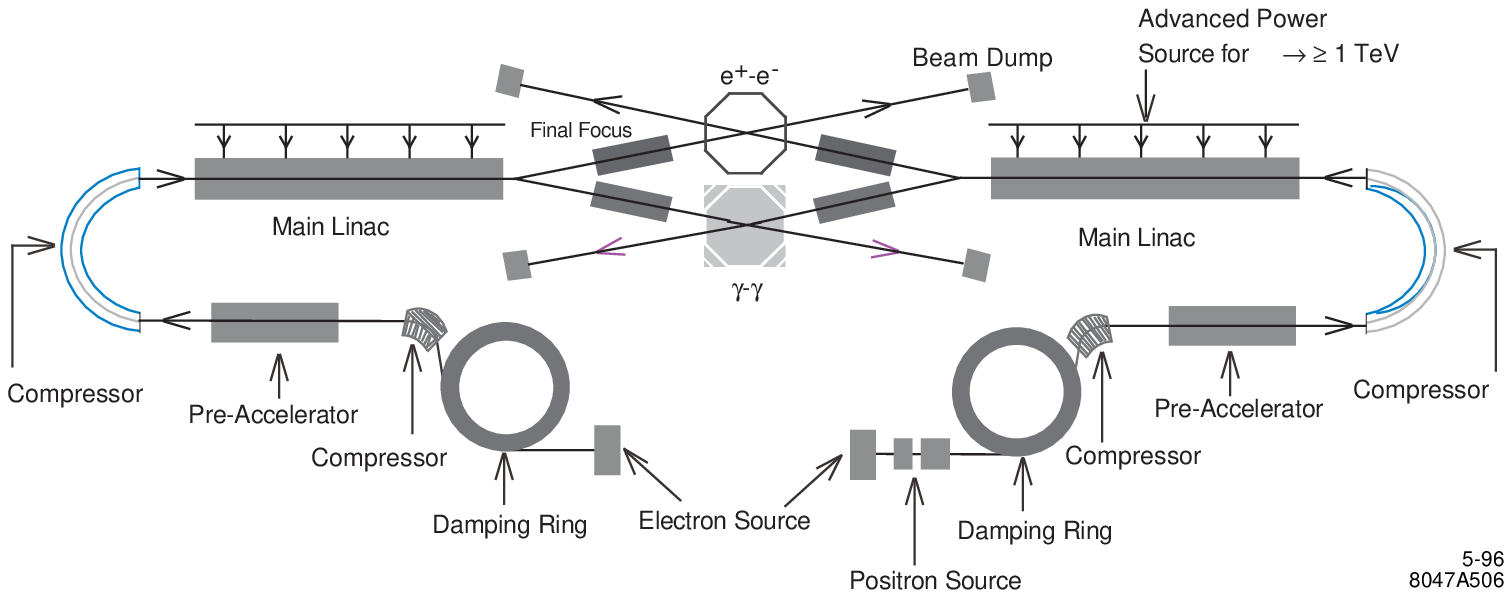}}
\captive{\label{fig:contr_01} 
 The general layout of a future linear collider, from 
 Ref. \protect\cite{ADO-9601}.
 }
\end{center}
\end{figure}

 There are several research programmes around the world which investigate
 different options on how to build a future linear collider. 
 The individual projects pursued today are, CLIC~\cite{BOS-9801},
 JLC~\cite{AKA-9701}, NLC~\cite{ADO-9601} and TESLA~\cite{BRI-9701}.
 This review only discusses the physics measurements which can be performed 
 at such a collider and it does not consider the arguments in favour or
 against the individual attempts and also not the prospects for the
 construction of a PLC.
 Only the general features of a generic linear collider,
 see Figure~\ref{fig:contr_01} for a sketch, 
 are addressed below.
%
%
 The linear collider is an extension of the existing 
 \epem colliders LEP and SLC. Table~\ref{tab:LC} shows the 
 improvements on several machine parameters which have to be achieved
 in order to arrive at a luminosity of the order of
 10$^{34}/$cm$^2$s, which would lead to an integrated luminosity 
 of about 100 fb$^{-1}$ per year of operation.
 \par
\begin{figure}[htb]
\begin{center}
{\includegraphics[width=0.8\linewidth]{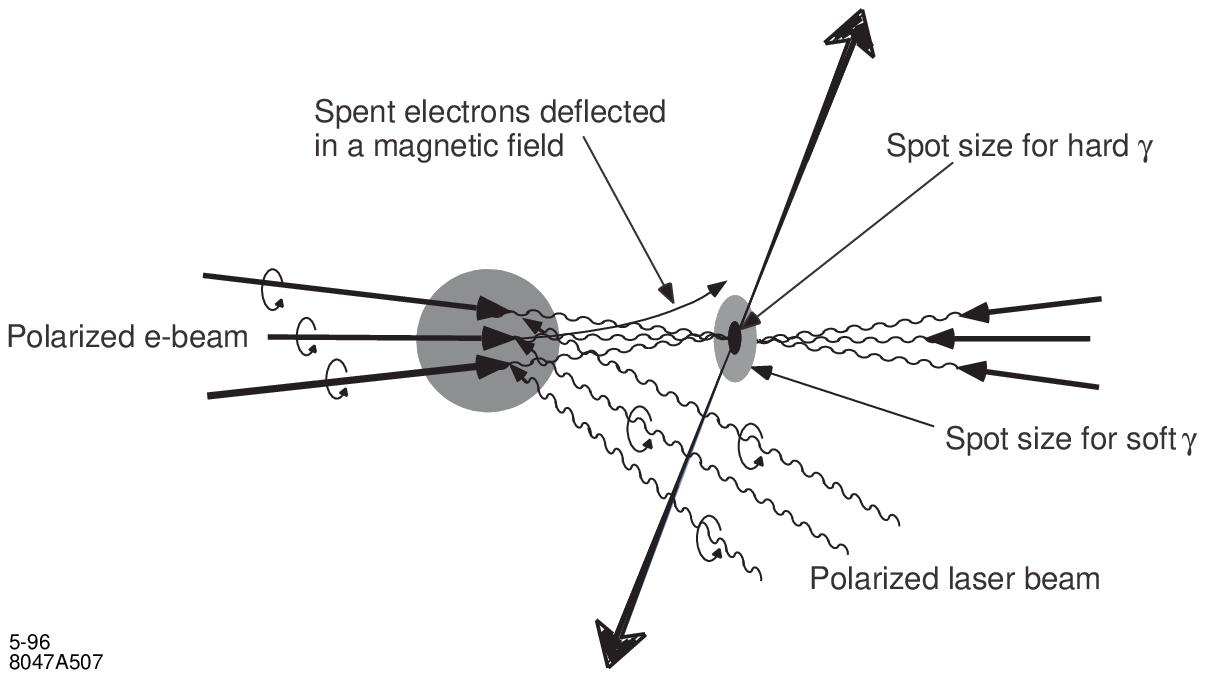}}
\captive{\label{fig:contr_02} 
 A sketch of the creation of the photon beam by Compton backscattering 
 of laser photons off the beam electrons, from Ref. \protect\cite{ADO-9601}.
 }
\end{center}
\end{figure}

 
%
\renewcommand{\arraystretch}{0.92}
\begin{table}[htbp]\begin{center}\begin{tabular}{|rr|cc|c|} 
\hline   
              &                       &  LEP     
                                      &  SLC 
                                      &  TESLA     \\
\hline
%
 total length & [km]                  &  26.7
                                      &  4 
                                      & 33         \\
 gradient     & [MV/m]                &  6
                                      &  10
                                      &  25         \\
 beam size  $\sigma_x/
 \sigma_y$    & [$\mu$m/$\mu$m]       &  110 / 5 
                                      &  1.4 / 0.5
                                      &  0.845/0.019 \\
 electron energy& [GeV]               &  100
                                      &  50
                                      &  250        \\
 luminosity   & [10$^{31}/$cm$^2$s]   &  7.4 
                                      &  0.1
                                      &  5000-10000 \\
 \lumi        & [1/pb y]              &  200
                                      &  15
                                      &  20000      \\\hline
\end{tabular}  
\captive{\label{tab:LC}
 Some approximate values of parameters of the present LEP and SLC colliders and
 goals for a future linear collider of the TESLA design.
 }
\end{center}\end{table}
%
 For several reactions the cross sections for incoming 
 photons are larger than for incoming electrons of the same energy,
 see Section~\ref{sec:selec}, and some reactions
 like e.g. the very important process $\GG\rightarrow H$
 are only possible in reactions of high energetic photons.
 Therefore there is a strong interest in the construction of a PLC,
 which would be an ideal source of high energetic photons.
 In order to build a PLC several constrains have to be imposed 
 on the machine and the interaction region of a \GG collider, shown 
 schematically in Figure~\ref{fig:contr_01}, has to be 
 designed differently than an interaction region for the \epem collider.
 The general procedure to produce a beam of high energetic photons from
 the electron beam by means of the Compton backscattering process
 is shown in Figure~\ref{fig:contr_02}.
 \par
%
\begin{figure}[htb]
\begin{center}
{\includegraphics[width=0.42\linewidth]{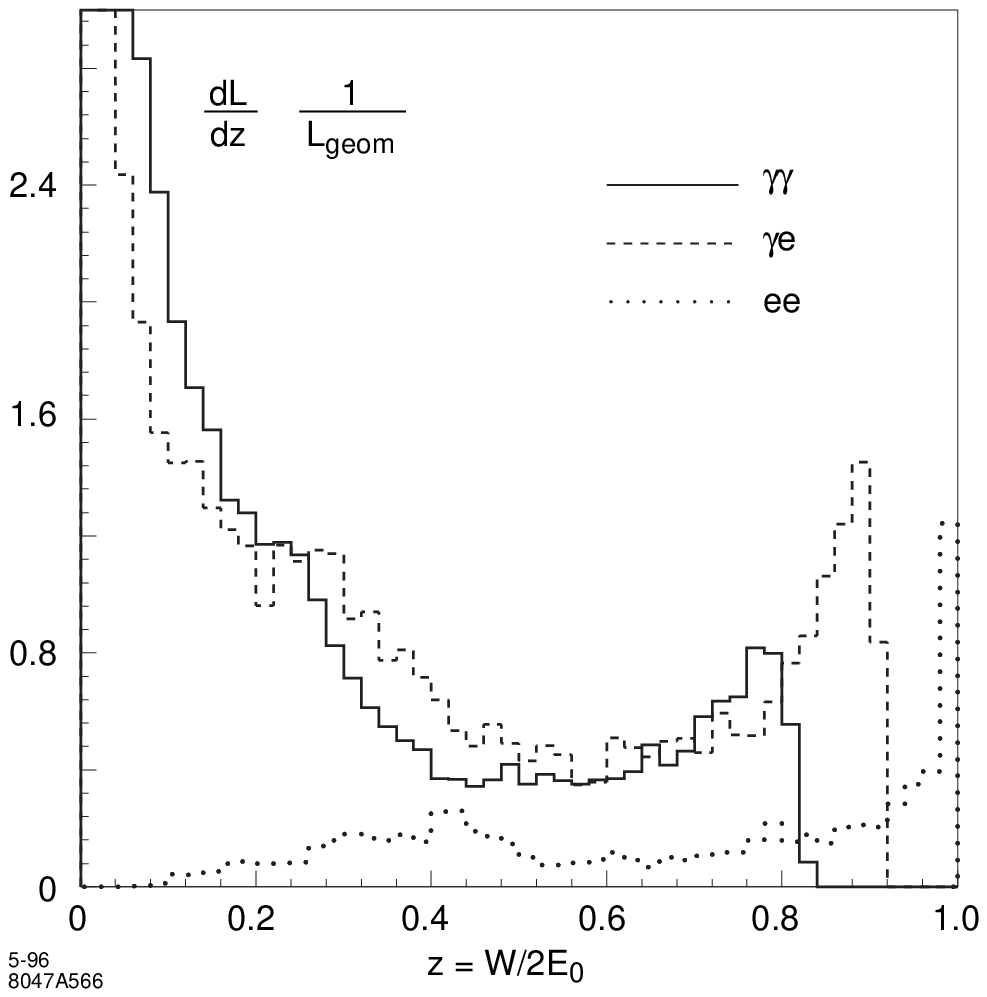}}
{\includegraphics[width=0.40\linewidth]{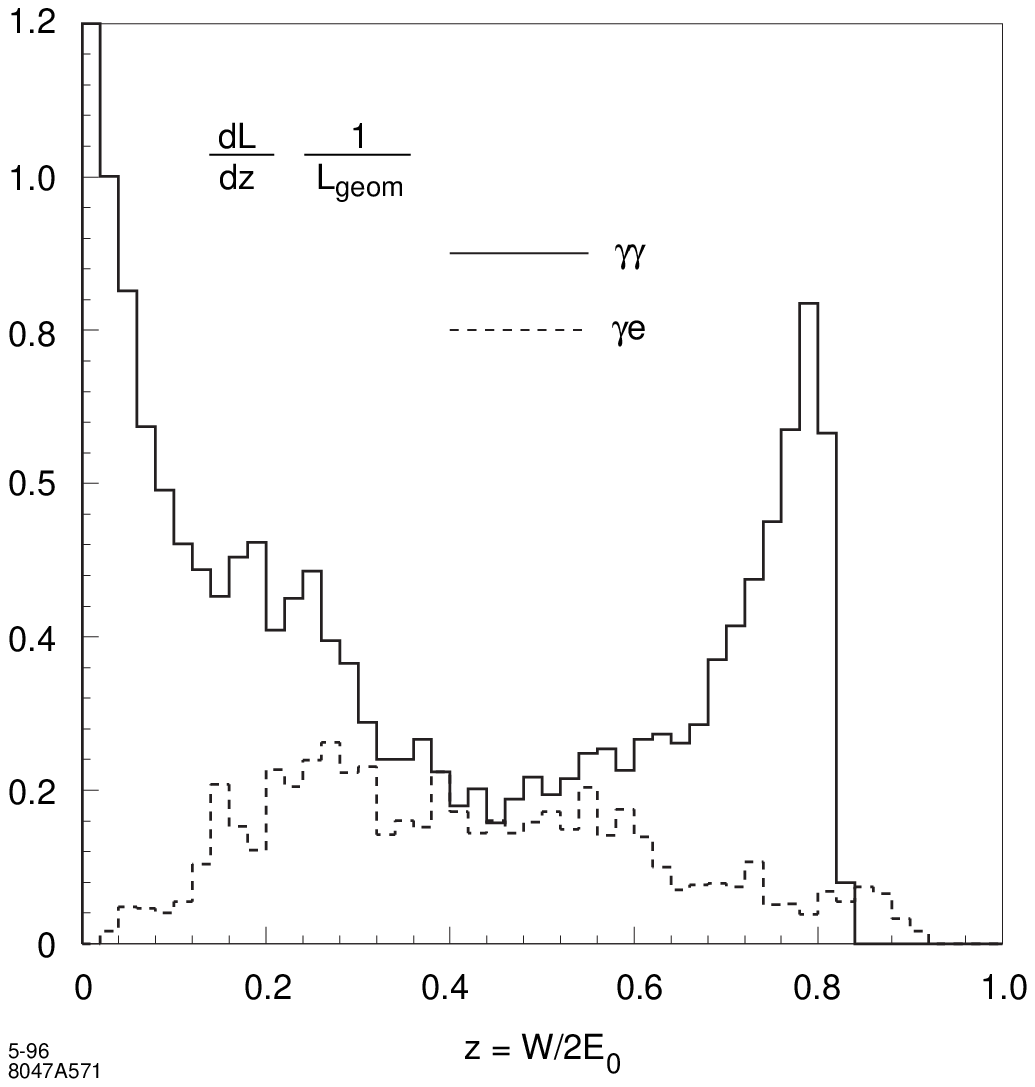}}
{\includegraphics[width=0.40\linewidth]{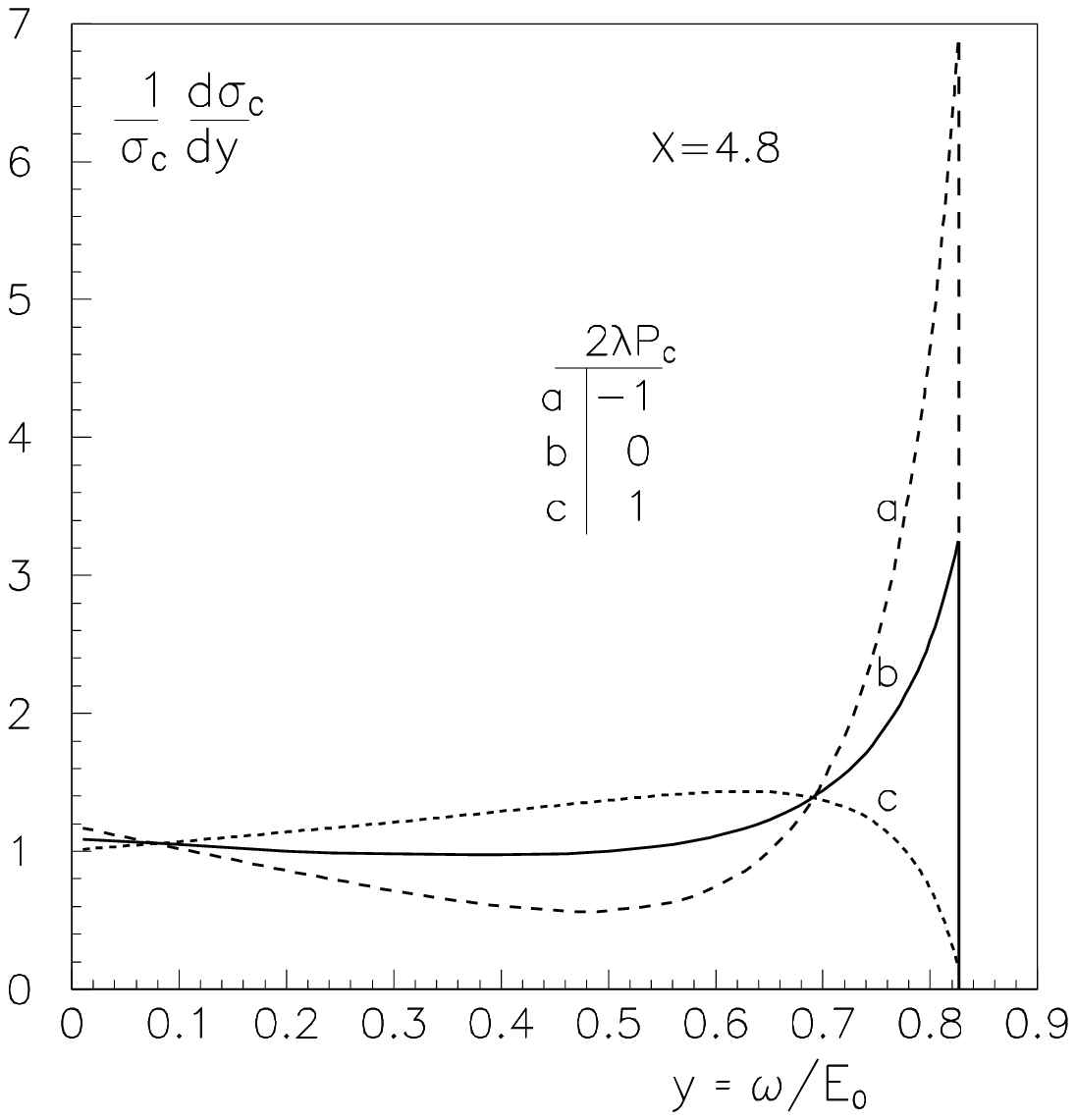}}
{\includegraphics[width=0.40\linewidth]{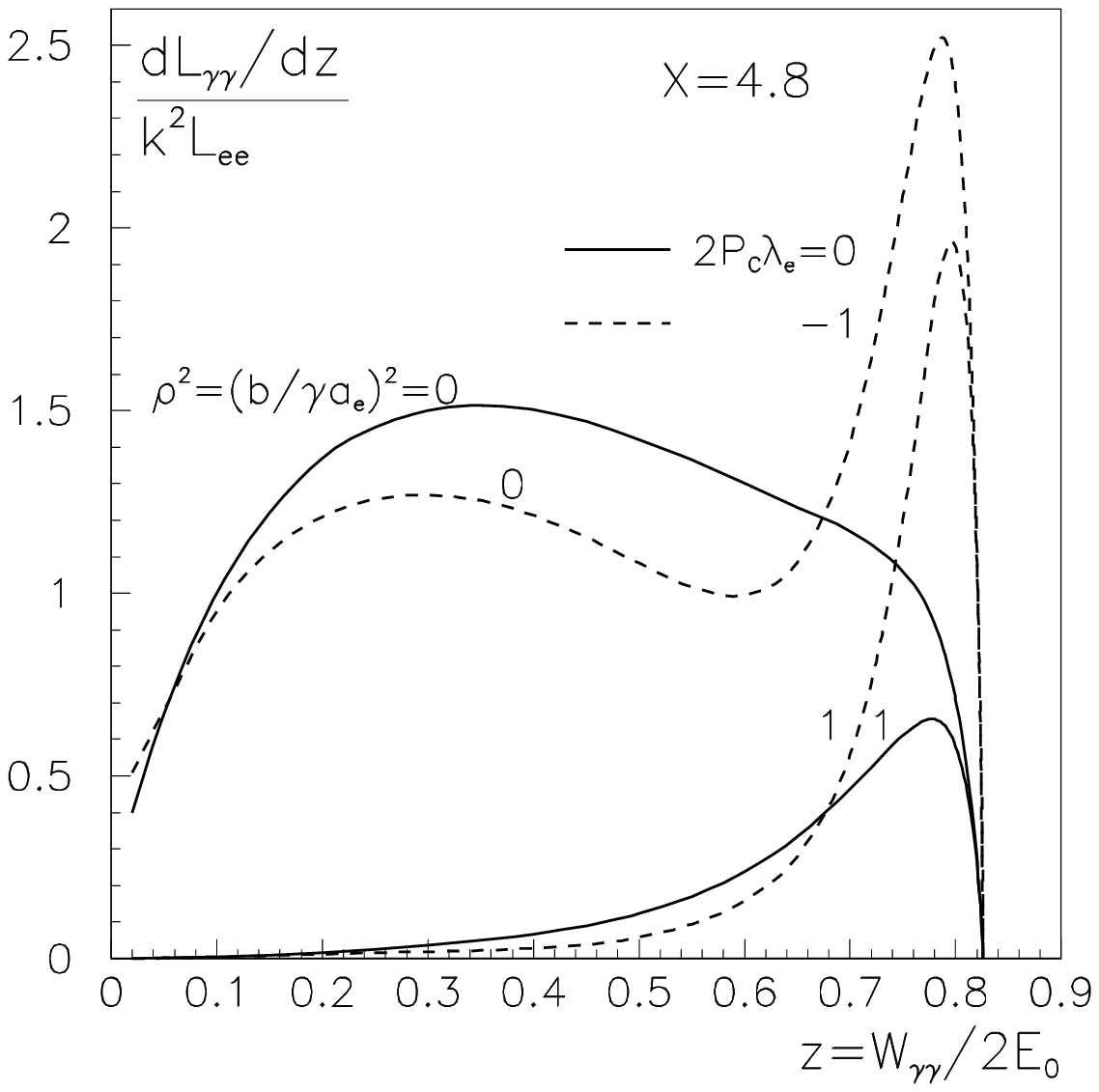}}
\begin{picture}(0,0)
\put(-220,350){\bf(a)}
\put( -30,350){\bf(b)}
\put(-300,160){\bf(c)}
\put(-120,160){\bf(d)}
\end{picture}
\captive{\label{fig:contr_03} 
 Some expected distributions for critical parameters of a PLC, 
 from Ref. \protect\cite{TEL-9701,ADO-9601}.
 Shown are 
 (a) the luminosity spectrum without deflection of the spent beam
     for a vertical offset of the beam of $0.75\sigma_y$ and a distance of
     $b=0.5$~cm,
 (b) the luminosity spectrum with deflection of the spent beam
     with $B=1$~T and a distance of $b=0.78$~cm,
 (c) the energy spectrum of the Compton scattered photons for fixed $x$ 
     and different polarisations of the laser photons and the 
     beam electrons as a function of the photon energy divided by the
     energy of the beam electrons,
 and
 (d) the luminosity spectrum as a function of the photon-photon
     centre-of-mass energy scaled by the \epem centre-of-mass energy
     for different polarisations and distances between the conversion 
     point and the interaction point.
 }
\end{center}
\end{figure}

%
 The photons are produced by a high intensity laser and brought into collision
 with the electron beams at distances of about 0.1$-$1 cm from the 
 interaction point. The photons are scattered into a small cone around
 the initial electron direction and receive 
 a large fraction of the electron energy.
 The opening angle is such that the high energetic photons produce 
 a smaller beam spot than the softer ones.
 If the Compton scattering is done on one side (two sides)
 than the collider runs in the $\eG$ ($\GG$ mode).
 For a \GG collider there will always be remaining \eG and  
 \ee luminosities, Figure~\ref{fig:contr_03}(a), 
 because some of the leftover electrons of the so called spent 
 beams will reach the interaction region. 
 The remaining \eG and \ee luminosities
 can considerably be reduced by magnetic deflection of the 
 spent beam, Figure~\ref{fig:contr_03}(b).
 \par
%
%
 The critical parameters of a PLC are the achievable 
 \GG luminosity,
 the energy spectrum of the Compton scattered photons, 
 the resulting polarisation of the photon beam, 
 the background produced at the interaction region,
 and the remaining
 \eG and \ee luminosities.
 Within limits several of these parameters can be chosen at 
 will~\cite{TEL-9701,TEL-9801} by changing the 
 distance along the beam line between of the production of the backscattered
 photons and the interaction region, $b$,
 by selecting the polarisation of the laser beam, \Pc, and
 the polarisation of electron beam, \lame,
 and by magnetic reflection of the spent beam, as
 illustrated in Figure~\ref{fig:contr_03}(c,d).
 The most peaked energy distribution is achieved for $2\Pc\lame=-1$ and
 $b\rightarrow 0$. Here $k=N_\gamma/N_{\rm e}$
 is the fraction of electrons that can be converted into photons and
 $a$ is the r.m.s radius of the electron beam at the interaction point.
 The symbols $\omega_0$ and $E_0$ denote the energies of the laser photons and
 beam electrons, respectively and $x=\frac{E_0\omega_0}{m^2c^4}$, with
 $m$ being the mass of the electron.
 The geometrical luminosity, ${\rm L}_{{\rm ee}} = {\rm L}_{{\rm geom}}$,
 is defined as ${\rm L}_{{\rm geom}} = N^2f/(4\pi\sigma_x\sigma_y)$,
 where $N$ is the number of particles in the beam, 
 $f$ is the repetition frequency, and $\sigma_x$ and $\sigma_y$ are the
 transverse beam sizes at the interaction point.
 In summary a typical distribution of \GG and \eG luminosity as a function
 of the invariant mass peaks at the maximum reachable 
 invariant mass of around $0.8\sqrt{s_{\ee}}$ 
 with widths of $\dWGG\approx 0.15$ for \GG, and $\dWeG\approx 0.05$ for 
 \eG collisions~\cite{TEL-9801}.
 \par
%
\begin{figure}[htb]
\begin{center}
{\includegraphics[width=0.4\linewidth]{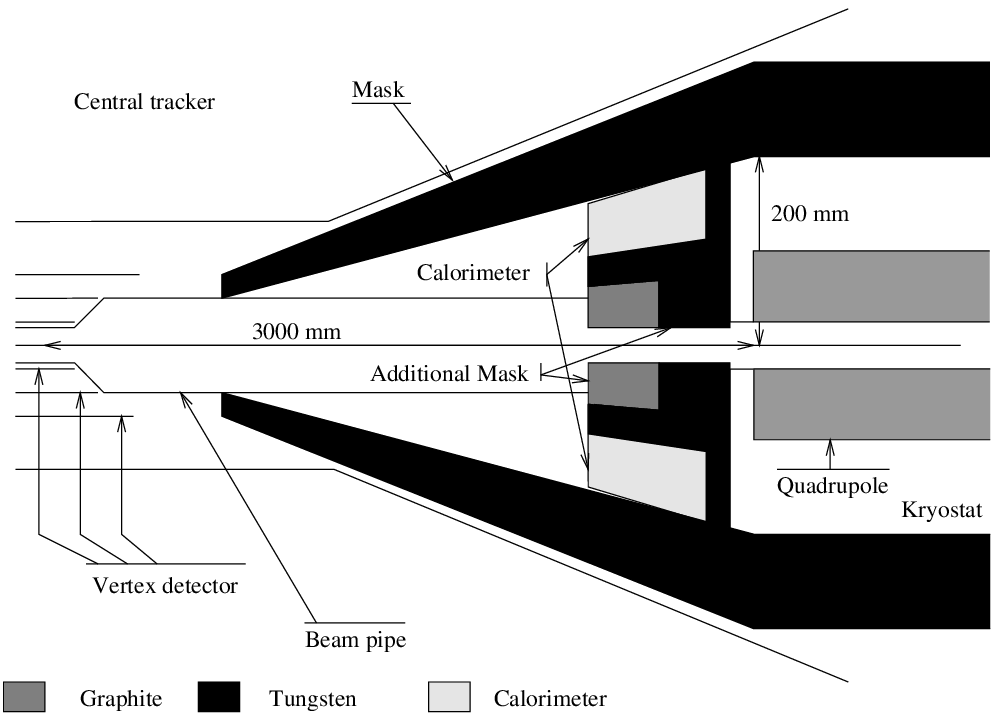}}
\captive{\label{fig:contr_04} 
 A sketch of the proposed mask for the TESLA design 
 to protect the detector from the background, from Ref.
 \protect\cite{SCU-9801}.
 }
\end{center}
\end{figure}
%
 The linear collider, even when running in optimal conditions will produce
 a huge amount of background where many particles are produced especially
 in the forward regions of the detector. In order to cope with this
 background the detector has to be shielded with a massive mask as shown
 e.g. for the TESLA design in Figure~\ref{fig:contr_04}.
 \par
%
\begin{figure}[htb]
\begin{center}
{\includegraphics[width=0.40\linewidth]{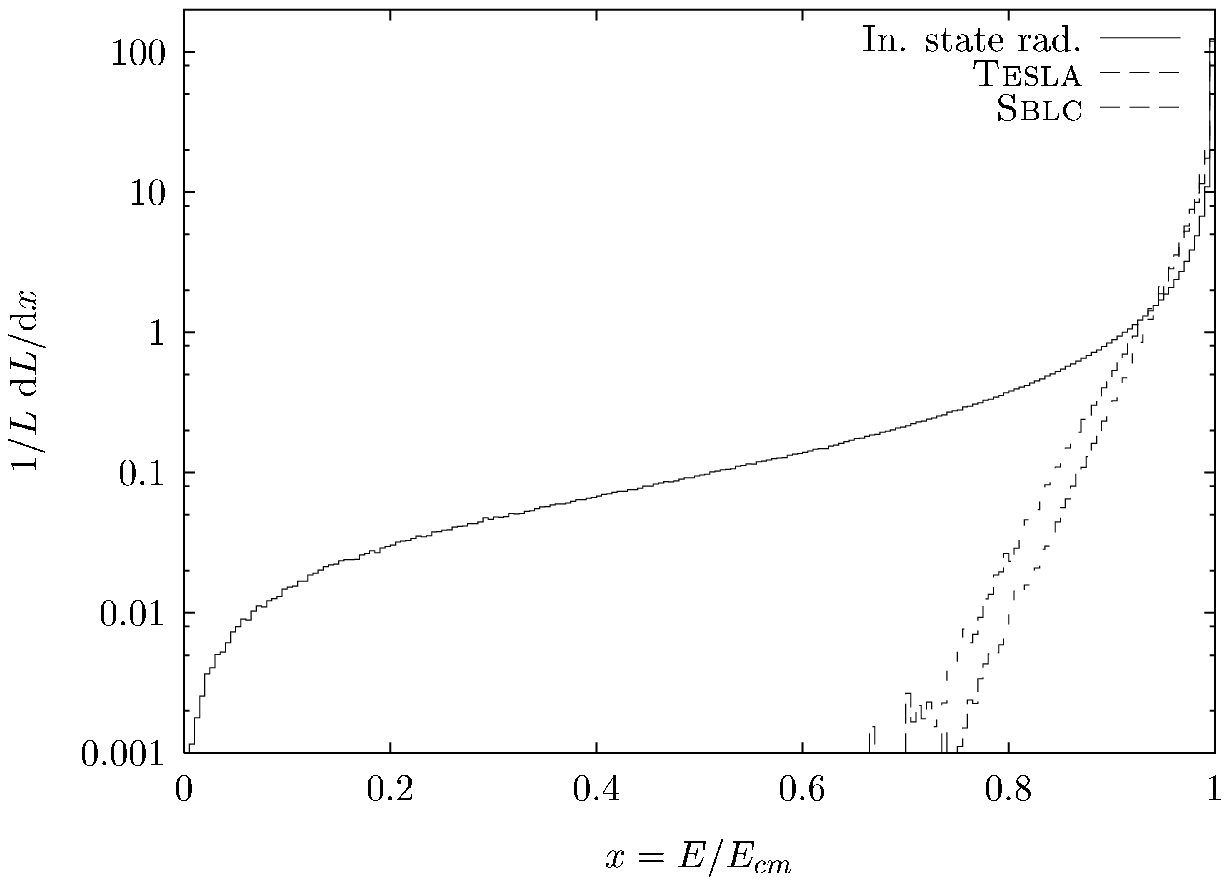}}
{\includegraphics[width=0.43\linewidth]{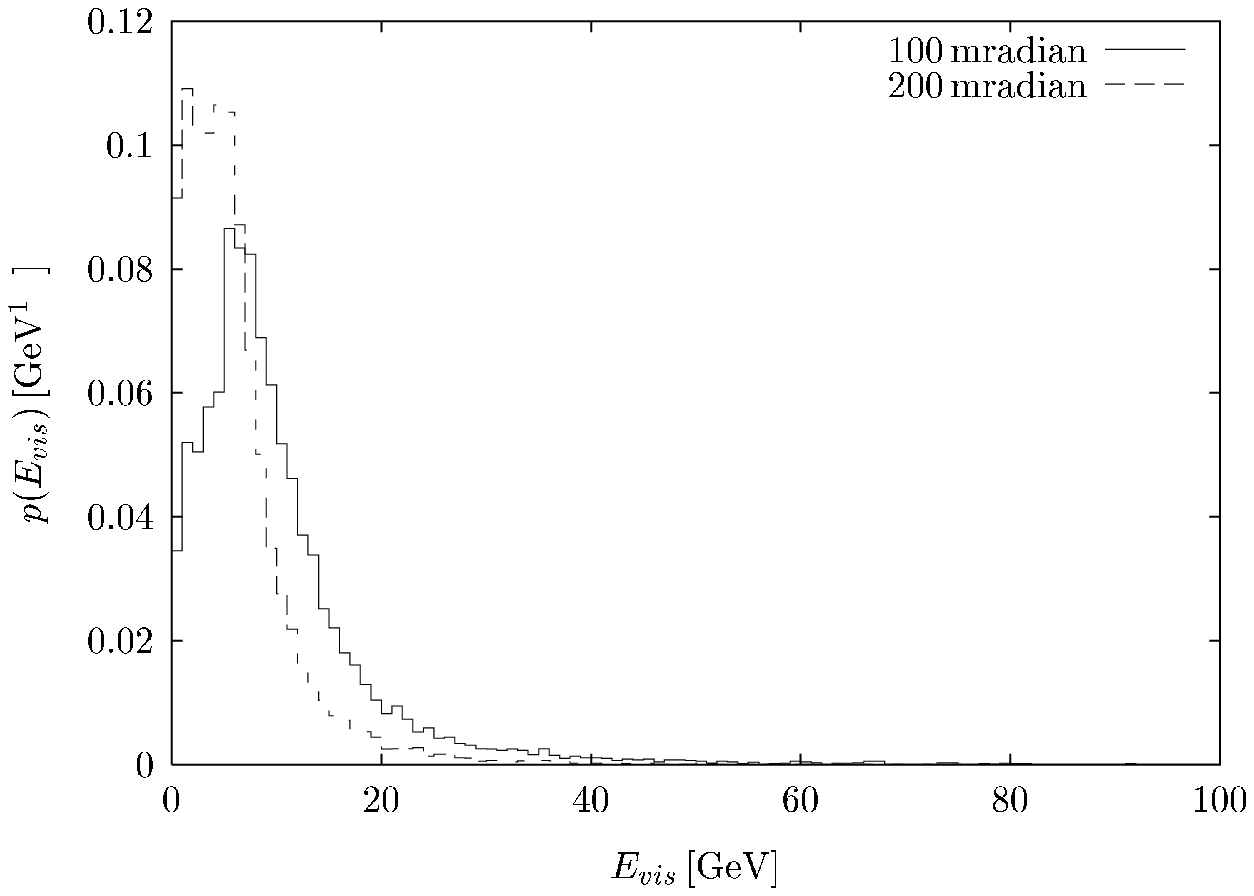}}
\begin{picture}(0,0)
\put(-230,35){\bf(a)}
\put( -35,35){\bf(b)}
\end{picture}
\captive{\label{fig:contr_05} 
 Some features of the expected background, from Ref. \protect\cite{SCU-9801}.
 Shown are 
 (a) the luminosity spectrum at the interaction point 
     as a function of the electron energy scaled by the nominal energy 
     of the beam electrons, where the energy losses shown are
     due to initial state radiation and beamstrahlung
 and
 (b) the distribution of the visible energy of hadronic events for two 
     different acceptance boundaries for the detection of hadrons
     of $100$~mrad and $200$~mrad.
 }
\end{center}
\end{figure}
%
 Detailed background studies for the linear collider were performed.
 The background sources are synchrotron radiation 
 in the last doublet of quadrupole magnets
 and the final focus system, muons accompanying 
 the beams, beamstrahlung (photon radiation of electrons of one beam
 in the strong field of the electrons of the other beam) which 
 will mainly lead to \epem pair creation, hadronic
 background and for a PLC also losses of the spent beams.
 The photon radiation will lead to a significant energy smearing 
 for the electrons of the beams. 
 The expected energy spectrum of the electrons at the interaction
 region is shown in Figure~\ref{fig:contr_05}(a).
 For the \epem mode the background simulation~\cite{SCU-9801}
 showed that the amount of background expected per bunch
 crossing for the TESLA design is about $10^5$ \epem pairs with a total 
 energy of $1.5\cdot10^5$ \gev and about 0.13 events of the type
 $\GG\rightarrow {\rm hadrons}$ for hadronic masses $W_{{\rm had}}>5$ \gev
 with an average visible energy of 
 $\langle E_{{\rm vis}}\rangle = 10$ \gev, see Figure~\ref{fig:contr_05}(b).
 The prospects of two photon physics measurements have to be discussed
 in the context of this expected machine parameter dependent
 'soft' underlying background.
%
%
\section{Some selected physics topics}
\label{sec:selec}
\begin{figure}[htb]
\begin{center}
{\includegraphics[width=0.7\linewidth]{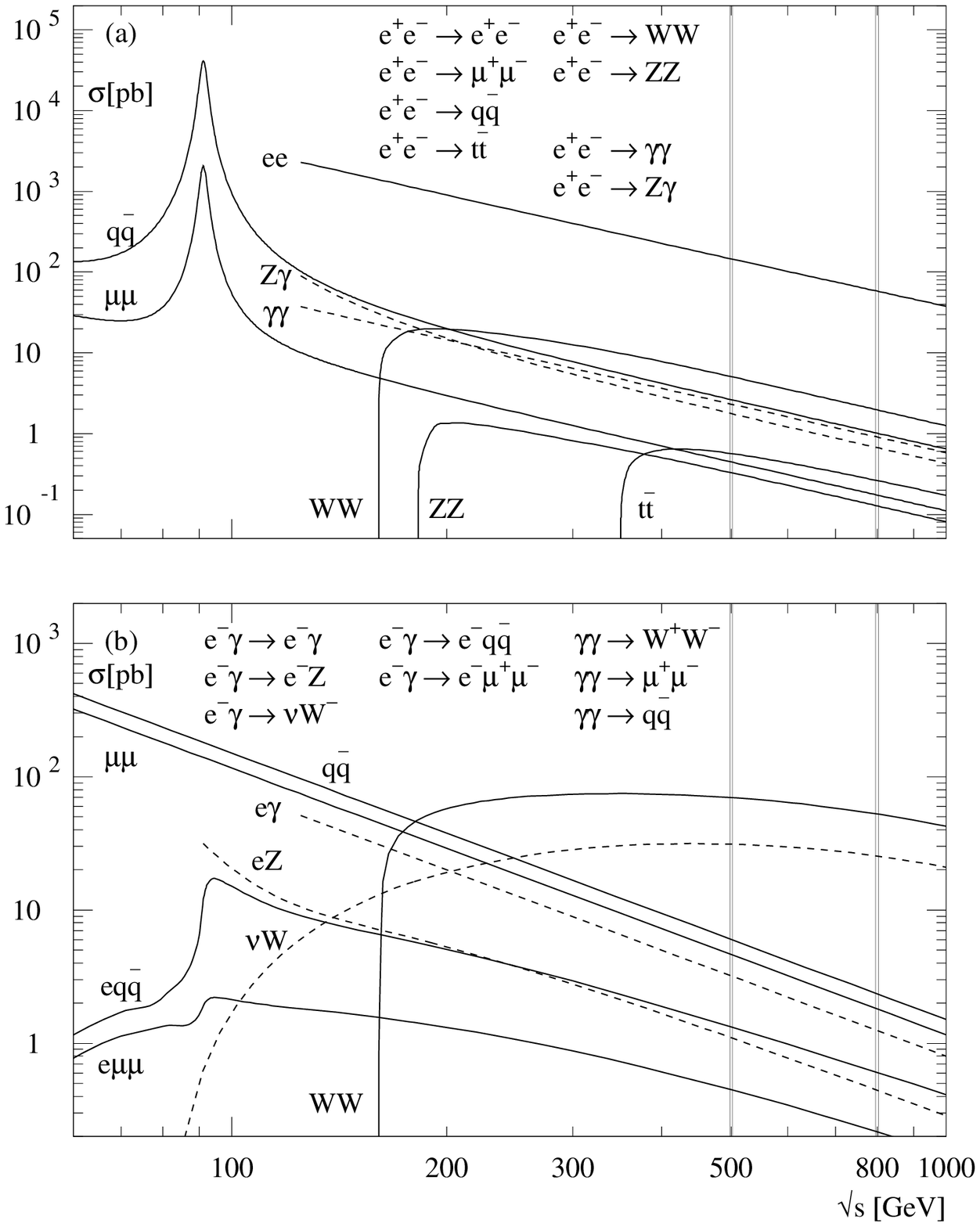}}
\captive{\label{fig:contr_06} 
 The expected cross sections as a function of the 
 centre-of mass energies, from Ref. \protect\cite{ACC-9801}.
 Shown are 
 (a) some \epem reactions
 and 
 (b) \emga and \GG reactions, 
 all as functions of the respective centre-of mass energy $\sqrt{s_{\ee}}$, 
 $\sqrt{s_{\emga}}$ or $\sqrt{s_{\GG}}$.
 }
\end{center}
\end{figure}

 The main processes that will be studied at a future linear collider 
 are \epem annihilation reactions. 
 Figure~\ref{fig:contr_06} shows the cross sections for those reactions 
 together with the cross sections for \emga and \GG reactions
 as a function of the respective centre-of-mass energy~\cite{ACC-9801}.
 The calculations are performed for a restricted range in polar 
 angle, $\theta$, for the outgoing particles or partons from the 
 hard sub-process, $10 < \theta <170$ deg. 
 In addition the invariant masses of the \mumu and \qqbar final states in
 the inelastic Compton processes are restricted to $M_{\mumu,\qqbar}>50$ \gev.
 \par
%
%
 For an integrated luminosity of about 20 fb$^{-1}$ per year of operation
 the expected event rate for a cross section of $1$~pb is 20000 events.
 It can be seen that e.g. the available cross section for $W$ production
 is much higher in the \GG mode than in the \epem mode at
 $\sqrt{s_{\ee}} = \sqrt{s_{\GG}}$. 
 Given this it is clear that the scenario for two-photon physics will be 
 very much depending on whether a PLC can be build or not.
 \par
 The following topics will be discussed:
 \begin{Enumerate}
 \item The total hadronic photon-photon cross section
 \item Jet cross sections in photon-photon scattering
 \item The measurement of the photon structure function \ft
 \item The charm part \ftc of the structure function \ft
 \item The signature of the BFKL Pomeron in $\sigma_{\gamma^\star\gamma^\star}$
 \item The production of $W$ pairs
 \item The Higgs discovery potential of the process 
       $\GG\rightarrow H\rightarrow \bbbar$,
       the total decay width $\Gamma(H\rightarrow\GG)$ and the reaction
       $\GG\rightarrow H\rightarrow Z Z$
 \end{Enumerate}
 Several topics are not covered, amongst those are e.g.
 resonances, searches for new particles other than the Higgs boson,
 diffraction, the production of photon pairs, and much more.
 The considerations for the selected topics are based on 
 the energy spectrum of 
 the bremsstrahlung photons using the equivalent photon approximation, EPA, 
 on the energy spectrum of the beamstrahlungs photons
 which strongly depends on the machine parameters
 and
 on the energy spectrum of the Compton scattered photons for specific 
 configurations of a PLC. 
 \par
%
\begin{figure}[htb]
\begin{center}
{\includegraphics[width=0.42\linewidth]{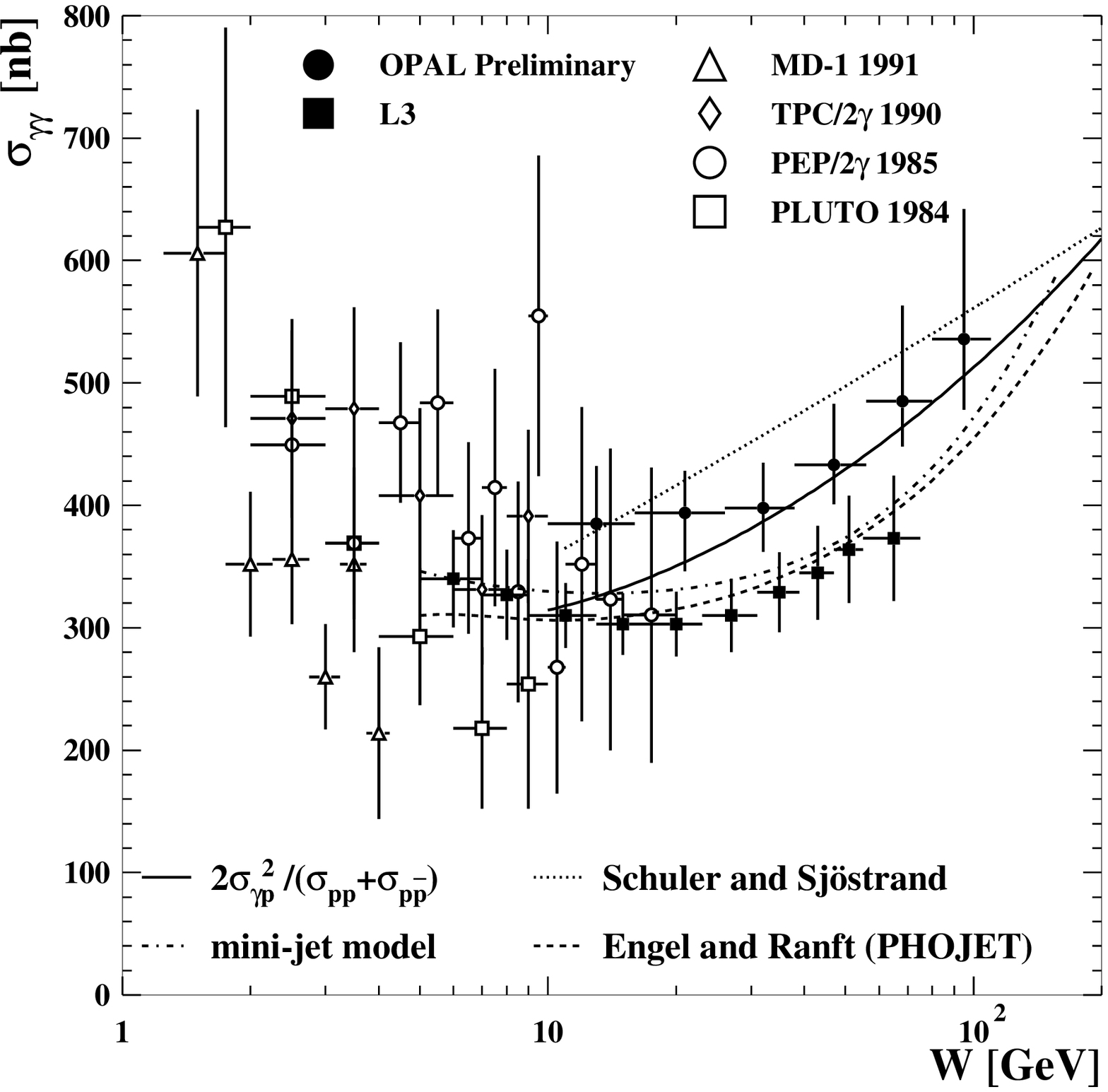}}
{\includegraphics[width=0.45\linewidth]{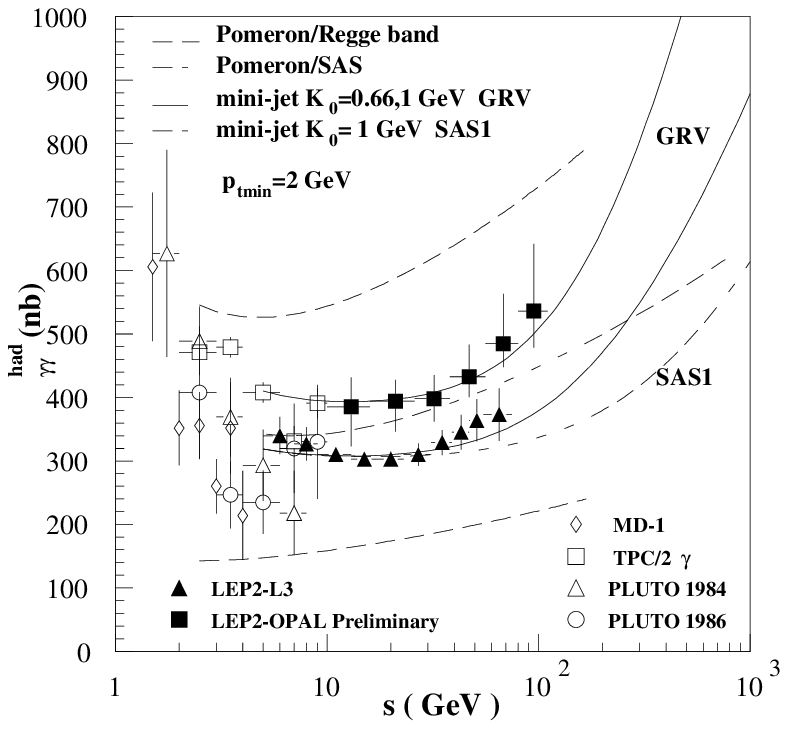}}
\begin{picture}(0,0)
\put(-255,60){\bf(a)}
\put( -50,60){\bf(b)}
\end{picture}
\captive{\label{fig:contr_07} 
 The total cross section measurements and theoretical expectations.
 Shown are
 (a) the LEP measurements together with the older data compared
     to several theoretical predictions, from Ref. \protect\cite{WAC-9701},
 and
 (b) a comparison of the data with predictions from the eikonalised minijet
     model and Regge based models with 
     extrapolations to higher energies, from Ref. \protect\cite{COR-9801}.
 }
\end{center}
\end{figure}
%
 The total hadronic cross section \stogg for the reaction 
 $\gamma \gamma\rightarrow {\rm hadrons}$ is of special theoretical
 interest as it allows to test theories which try to consistently describe
 proton-antiproton, proton-proton, photon-proton and photon-photon 
 interactions~\cite{DON-9601-SCH-9701-ENG-9501-ENG-9601,COR-9801}.
 Only a slow rise as function of the photon-photon invariant mass is predicted
 for \stogg, which means that a very large lever arm is needed in order
 to disentangle the different slopes of the various predictions. 
 Compared to the pre LEP data there is already quite an improvement 
 on precision and lever arm from the LEP data, see 
 Figure~\ref{fig:contr_07}, and indeed a slow rise is seen in 
 the OPAL~\cite{WAC-9701} and L3~\cite{L3C-9704} data.
 This measurement can be extended to larger values
 of $W$ with a future linear collider, but it has to be kept in mind that
 the precision of the present LEP data is limited by the uncertainty of the 
 theoretical description of the observable and especially also the 
 unobservable hadronic final states as implemented in Monte Carlo models.
 \par
%
\begin{figure}[htb]
\begin{center}
{\includegraphics[width=0.40\linewidth]{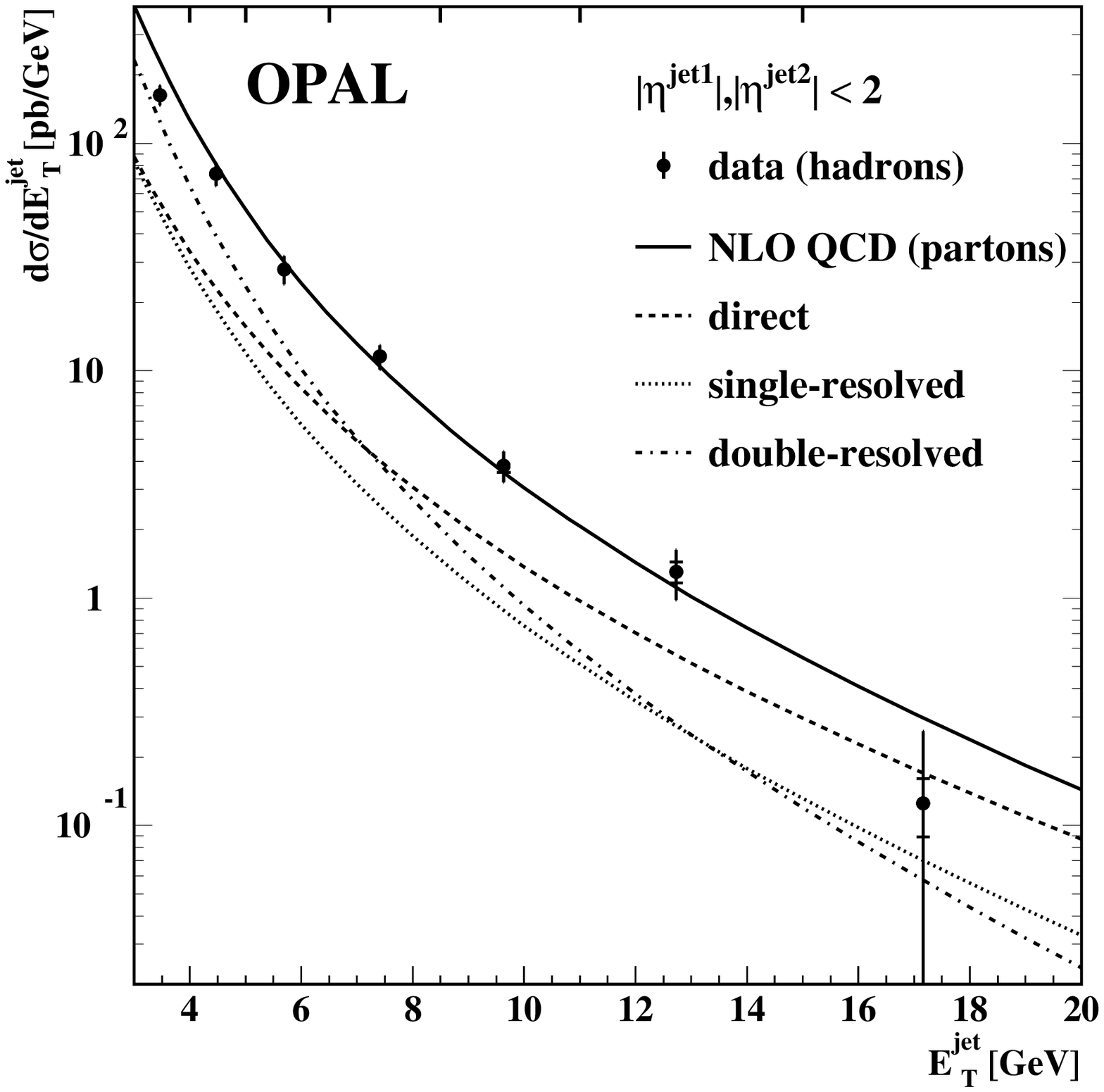}}
{\includegraphics[width=0.50\linewidth]{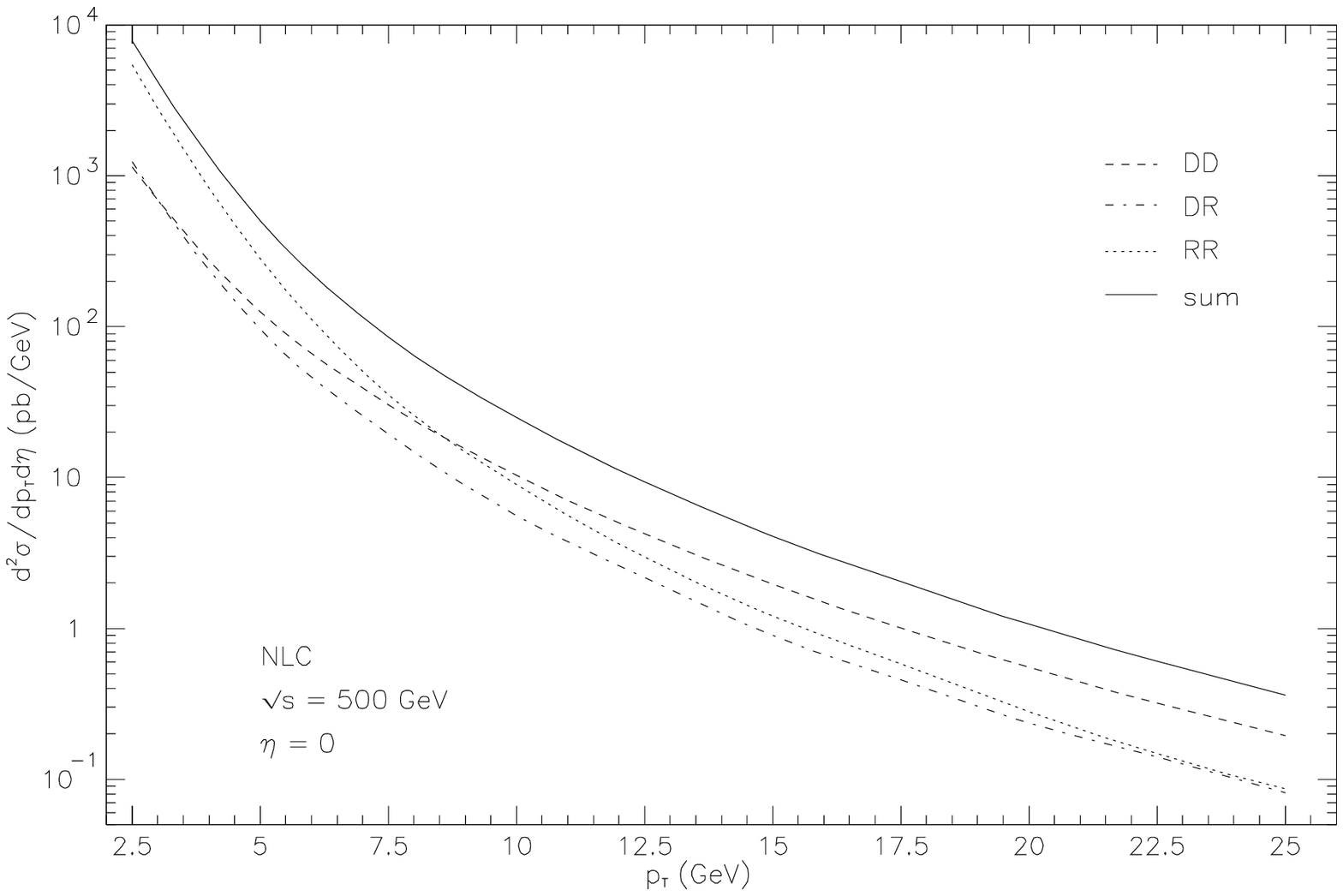}}
\begin{picture}(0,0)
\put(-385,30){\bf(a)}
\put(-210,30){\bf(b)}
\end{picture}
\captive{\label{fig:contr_08} 
 Measured inclusive jet cross sections \protect\cite{OPALPR250}
 and the accessible range for a future linear collider.
 Shown are 
 (a) the inclusive di-jet cross sections from OPAL compared to the 
     NLO predictions of Ref. \protect\cite{KLE-9601}
 and
 (b) the NLO predictions for a future linear collider,
     where D stands for direct and R for resolved.
 }
\end{center}
\end{figure}
%
 Looking more exclusively than the total cross section 
 NLO jet cross sections can be confronted with the data. 
 This has been done at 
 LEP~\cite{OPALPR250}, Figure~\ref{fig:contr_08}(a), and
 good agreement was found between the inclusive jet cross sections 
 observed on the hadronic level and the NLO calculations
 of Ref.~\cite{KLE-9601}.
 This measurement gives information on the relative amount of the 
 direct processes,
 in which the photon directly takes part in the hard interaction, 
 and the resolved processes, where either one or both photons resolve into 
 a partonic state and only one of the partons, either a quark or a gluon, 
 takes part in the hard interaction.
 These measurements can be extended, Figure~\ref{fig:contr_08}(b),
 to larger jet transverse momenta\cite{KLE-9601}.
 However it has to kept in mind that in this region the jet cross section
 is dominated by direct processes where not much information on the 
 internal structure of the photon can be obtained.
 This fact can be circumvented by using the di-jet sample and
 separating the direct and the resolved processes by measuring 
 the fraction of the photon 
 momentum, $x_\gamma$, which participates in the hard 
 interaction~\cite{OPALPR250}.
 Another important part is the region of low jet transverse momenta
 which is dominated by processes initiated by the gluons in the photon.
 Again this is a 
 theoretically difficult region because the jet transverse momentum 
 is the hard scale in the process which should not get too small in order
 for theoretical predictions to be reliable.
 \par
%
\begin{figure}[htb]
\begin{center}
{\includegraphics[width=0.5\linewidth]{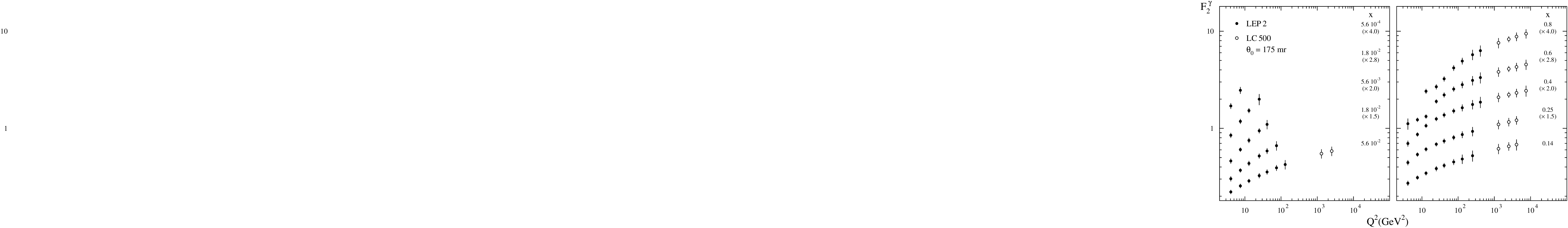}}\\
{\includegraphics[width=0.5\linewidth]{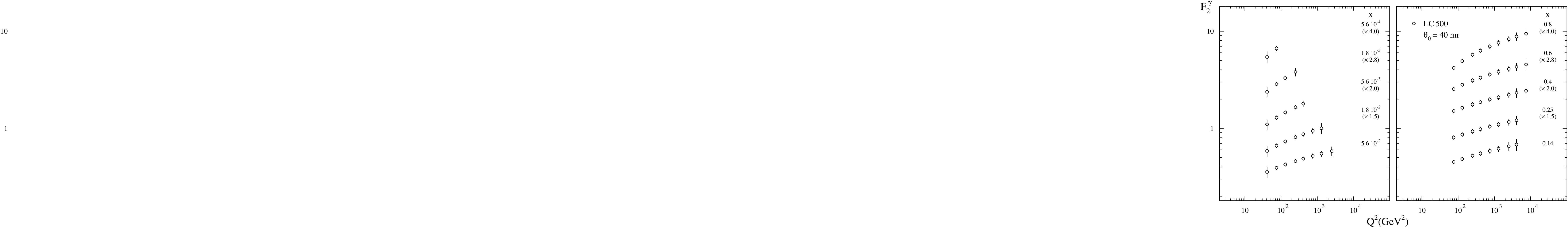}}
\begin{picture}(0,0)
\put(-110,210){\bf(a)}
\put(  30,210){\bf(b)}
\put(-110, 30){\bf(c)}
\put(  30, 30){\bf(d)}
\end{picture}
\captive{\label{fig:contr_09} 
 The prospects for structure function measurements at a future linear 
 collider, from Ref. \protect\cite{ACC-9801}.
 Shown are hypothetical LEP data and linear collider data for 
 different minimal detection angles of the scattered electrons, \ttag.
 In 
 (a,b) $\ttag > 175$~mrad is assumed
 and 
 (c,d) is based on $\ttag > 40$~mrad.
 }
\end{center}
\end{figure}
%
 Structure function measurements are an active research project at LEP
 and the results cover the \qsq range from about 1.5 to 300 \gevsq and the 
 $x$ range from 0.001 to about 1. 
 Two main questions are addressed, the behaviour of the photon structure
 function \ft at low values of $x$ and the \qsq evolution of the structure
 function \ft at medium $x$, see Ref.~\cite{NIS-9702} for a review. 
 Both these topics can be studied at a future linear collider but 
 stringent requirements have to be imposed on the detector 
 design~\cite{ACC-9801}.
 The region of high \qsq and high $x$ can already be studied with an 
 electromagnetic calorimeter located outside the shielding mask
 covering polar angles of the tagged electrons of $\ttag > 175$ mrad
 which is able to detect electrons with energies 
 above 50$\%$ of the beam energy, Figure~\ref{fig:contr_09}(a,b).
 The errors shown in Figure~\ref{fig:contr_09} are the quadratic sum 
 of the statistical and the systematic components.
 The statistical error is calculated based on the GRV LO structure 
 function \ft~\cite{GLU-9201-GLU-9202} for an integrated luminosity of 
 10 fb$^{-1}$. The systematic error is assumed to be equal to the 
 statistical error but amounts to at least 5$\%$.
 Therefore the precision indicated in Figure~\ref{fig:contr_09} has to 
 be taken with care as the systematic errors shown do not reflect the 
 present level of precision of the LEP data.
 In order to achieve overlap in \qsq with the LEP data the electron 
 detection has to be possible down to 
 $\ttag > 40$~mrad, Figure~\ref{fig:contr_09}(c,d),
 which means the mask has to be instrumented, and the calorimeter has 
 to be able to detect
 electrons which carry 50$\%$ of the beam energy in the huge but flat
 background of electron pairs discussed in Section~\ref{sec:instru}.
 \par
\begin{figure}[htb]
\begin{center}
{\includegraphics[width=0.6\linewidth]{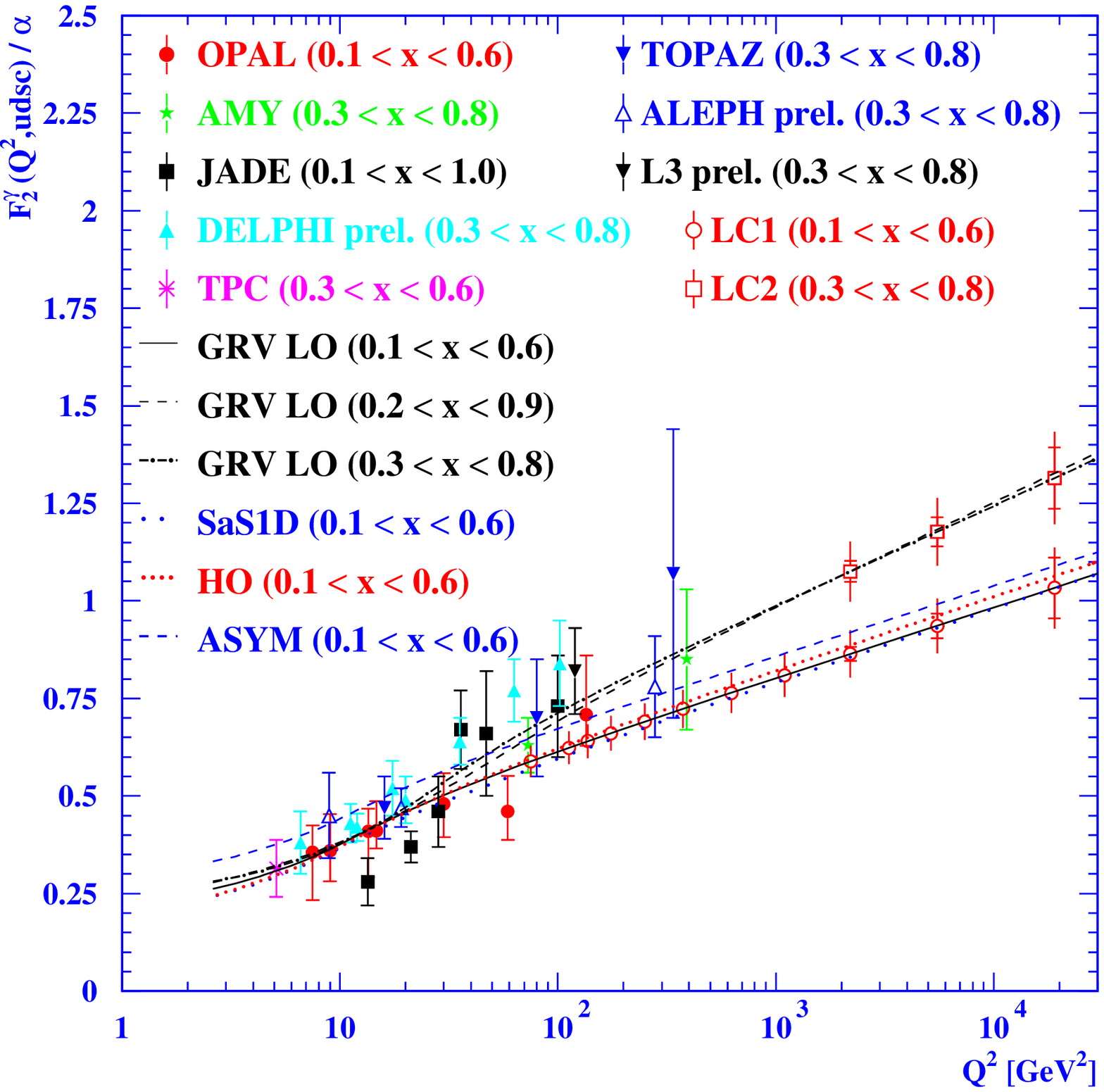}}
\captive{\label{fig:contr_10} 
 The measured \qsq evolution of \ft from Ref. \protect\cite{NIS-9702}
 extended by a preliminary measurement from L3 and by the prospects 
 of a measurement at a future linear collider.
 }
\end{center}
\end{figure}

 The measurement of the \qsq evolution of the structure function \ft 
 constitutes a fundamental test of QCD. The status of the measurements 
 as of today is reviewed in Figure~\ref{fig:contr_10}, which is 
 taken from Ref.~\cite{NIS-9702} and extended
 by adding the preliminary measurement from L3 at $\qsq=120$ \gevsq,
 see Ref.~\cite{FRE-9801}.
 The prospects of the extension of the measurement at a future linear 
 collider with $\sqrt{s}= 500$~\gev are shown for two scenarios.
 It is assumed that electrons can be tagged for energies $\etag/\eb>0.5$ and 
 for angles of $\ttag > 40$~mrad (LC1) and $\ttag >175$~mrad (LC2).
 The measured values are taken to be equal 
 to the prediction of the leading order GRV photon structure
 function \ft in the respective ranges in $x$, which are chosen to be 
 $0.1 < x < 0.6$ for LC1 and $0.3 < x < 0.8$ for LC2.
 The statistical errors of the hypothetical
 measurements are calculated from the number of events as predicted by 
 the HERWIG Monte Carlo~\cite{MAR-8801-KNO-8801-CAT-9101-ABB-9001-SEY-9201} 
 for the leading order GRV photon structure
 function \ft in bins of \qsq using the ranges in $x$ as indicated
 in Figure~\ref{fig:contr_10}. 
 The systematic error is assumed to be 6.7$\%$ and 
 to be independent of \qsq.
 This assumption is based on the systematic error of the 
 published LEP result with the highest \qsq from OPAL~\cite{OPALPR207}.
 The symmetrised value of the published systematic error 
 at $\qsq=135$~\gevsq is 13.4$\%$.
 It is assumed that this error can be improved by a factor of two.
 With this assumptions the error on the measurement is dominated
 by the systematic error up to the highest \qsq values.
 It is clear from Figure~\ref{fig:contr_10}
 that overlap in \qsq with the existing data can only be achieved if  
 electron detection with $\ttag > 40$~mrad is possible. For
 $\ttag > 175$~mrad sufficient statistics is only available for
 \qsq above around 1000 \gevsq.
 \par
 In summary with the data from the linear collider the measurement of 
 the \qsq evolution of the structure function \ft, 
 Figure~\ref{fig:contr_10}, can be extended to about $\qsq = 20000$ \gevsq
 and the behaviour of \ft at low values of $x$ can be investigated
 down to $x\approx 5\cdot 10^{-2}$ ($x\approx 5\cdot 10^{-4}$) for an 
 electron acceptance of $\ttag > 175$ mrad 
 ($\ttag > 40$ mrad)~\cite{ACC-9801}.
 \par
%
%
 For a PLC a completely new scenario for photon structure function
 measurements would be opened. For the first time measurements could be 
 performed with beams of high energetic photons of known energy with a rather
 small energy spread instead of measurements using
 the broad bremsstrahlungs spectrum of photons radiated by electrons.
 With this the measurements of the photon structure function would be on a
 similar ground than the measurement of the proton structure function at
 HERA.
 Another very important improvement for structure function measurements
 would be the detection of the electron that radiates the quasi-real photon
 and is scattered under almost zero angle. If this could be achieved the 
 precision of structure function measurements would significantly be improved,
 because $x$ could be calculated from the two detected electrons, 
 and therefore independently of the hadronic final state. 
 Given that the dominant systematic error 
 of the structure function measurement comes from the imperfect 
 description of the hadronic final state by the Monte Carlo models,
 this would be a important step to reduce
 the systematic error of structure function measurements.
 \par
%
\begin{figure}[htb]
\begin{center}
{\includegraphics[width=0.5\linewidth,angle=-90]{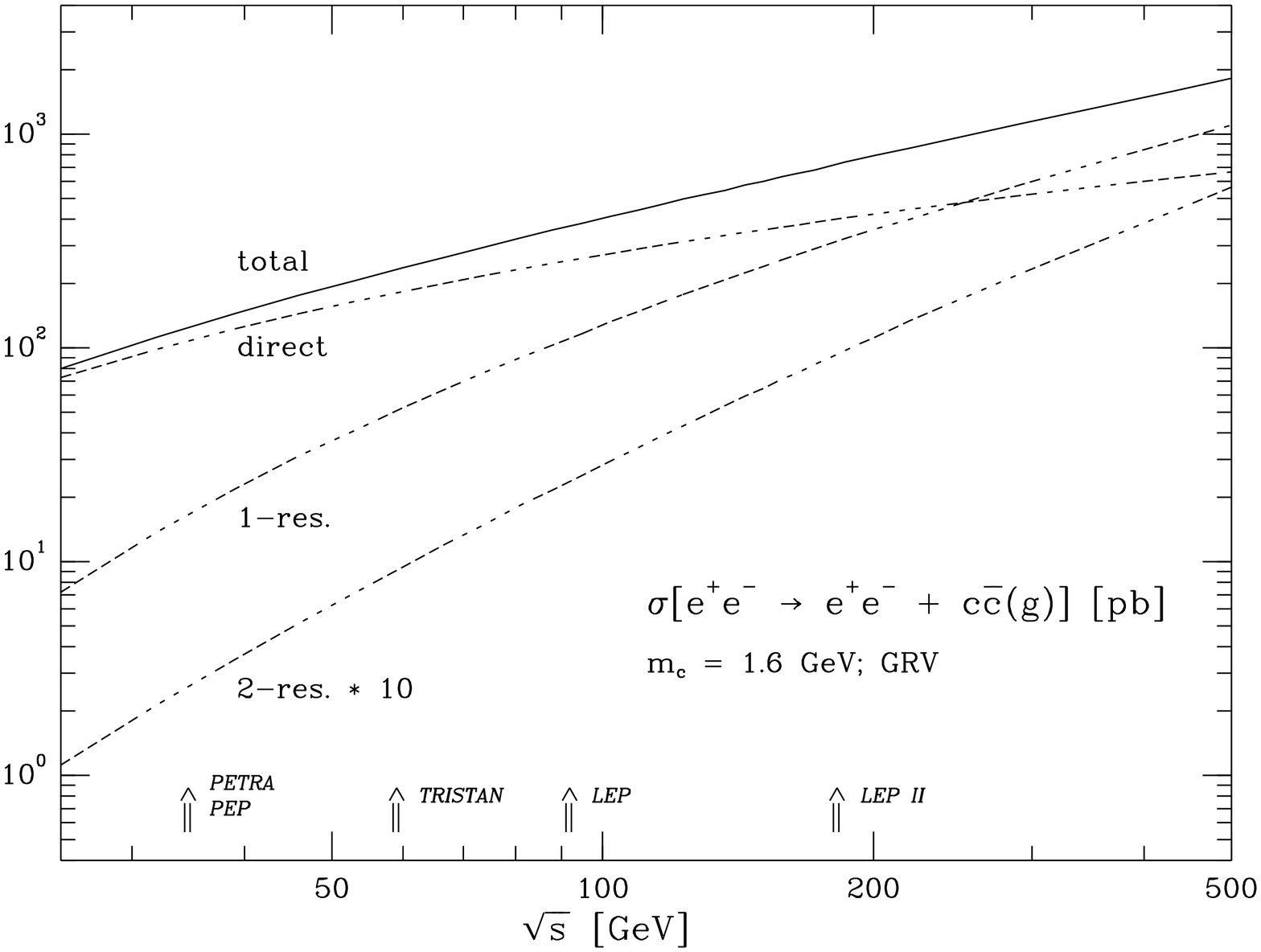}}
\captive{\label{fig:contr_11} 
 The charm cross section for photon-photon interactions, 
 from Ref. \protect\cite{DRE-9301}. Shown are the total cross section 
 (full line) and the individual contributions, direct (direct),
 single resolved (1-res) and double resolved (2-res).
 }
\end{center}
\end{figure}
%
 Pairs of charm quarks will be copiously produced at a future linear collider.
 Figure~\ref{fig:contr_11} shows the calculation from Ref.~\cite{DRE-9301} 
 for the production of \ccbar pairs in photon-photon scattering
 based on the bremsstrahlungs photons approximated by the EPA.
 The prediction is about $4\cdot10^7$ \ccbar events
 with $W_{{\rm min}} = 3.8$ GeV
 for an integrated luminosity of 20 fb$^{-1}$.
 This calculation takes into account the direct and the single-resolved 
 contribution in NLO, and the double-resolved contribution,
 which is much smaller, in LO.
 The mass of the charm quark is taken to be $\mc=1.6 $~GeV, 
 the renormalisation scale is set to $\mu = \sqrt{2}\mc$ and
 $\Lambda^{(4)}=340$~MeV in the $\overline{\rm MS}$ scheme.
 The cross section of the direct process is a pure QCD prediction 
 which only depends on the mass of the charm quark and on \al.
 It has been shown~\cite{CAC-9601} that a fair amount of these events 
 can be observed within the acceptance of the detector allowing to 
 test this pure QCD prediction.
 \par
%
\begin{figure}[htb]
\begin{center}
{\includegraphics[width=0.5\linewidth]{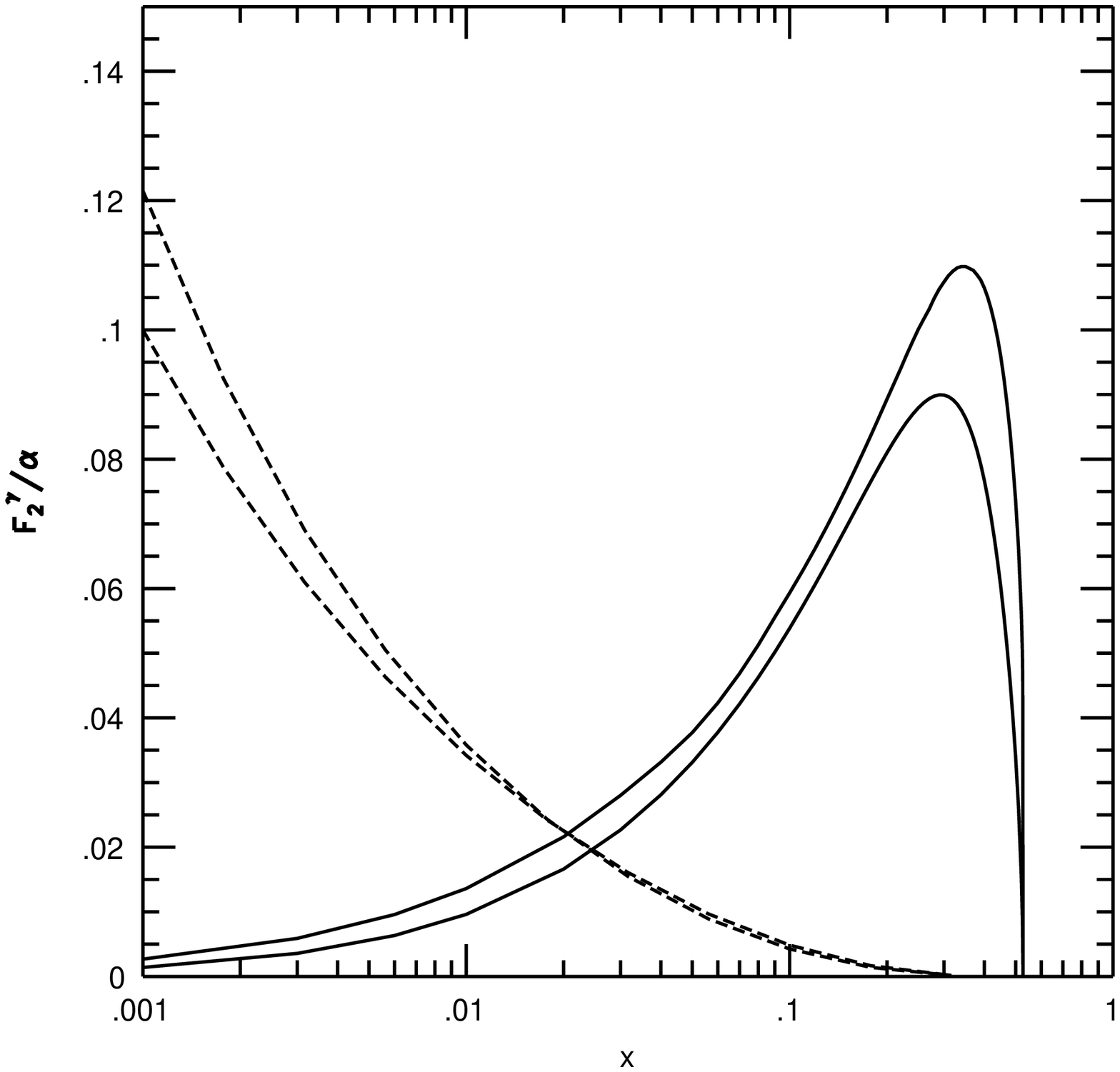}}
\captive{\label{fig:contr_12} 
 The expected charm contribution to \ft at $\qsq = 10$~\gevsq in LO and 
 NLO, from Ref. \protect\cite{LAE-9602}. 
 The dashed lines denote the hadronic and the full lines the point-like
 contributions to \ftc.
 The lower solid line is the LO and the upper solid line the NLO 
 prediction. At $x=0.001$ the lower dashed line is the NLO and the upper 
 dashed line is the LO prediction. 
 }
\end{center}
\end{figure}
%
 Also in the case of deep inelastic electron-photon scattering 
 charm quark pairs are frequently produced~\cite{LAE-9602}.
 The contribution of the individual quark species to the structure
 function \ft is proportional to the quark charge squared, which means that
 at high invariant masses the charm contribution to \ft should be 
 almost as large as the contribution from up quarks. 
 Due to the low efficiency for charm tagging the charm quark contribution
 to \ft has never been measured.
 The structure function \ftc receives two contributions which
 are clearly separated in $x$, Figure~\ref{fig:contr_12}. At low values 
 of $x$ the hadronic contribution dominates, whereas the point-like 
 contribution is concentrated at high values of $x$.
 The NLO corrections are rather small, see Figure~\ref{fig:contr_12}, 
 indicating a good stability of the perturbative QCD prediction.
 The hadronic contribution is directly proportional to the gluon density in 
 the photon and due to the large mass of the charm quark 
 the point-like contribution is a pure QCD prediction which is
 unambiguously defined to NLO~\cite{LAE-9602}.
 Given this a simultaneous measurement of the gluon density in the photon
 and of a pure QCD prediction is possible using the production of 
 charm quark pairs in deep inelastic electron-photon scattering.
 The calculations from Ref.~\cite{LAE-9602} show that for
 $\etag/\eb>0.5$, $\ttag>40$ mrad, $\mc=1.5$ \gev, $\mu=Q$
 and a charm tagging efficiency of 1-2$\%$ several thousand events
 could be seen for an integrated luminosity of 20 fb$^{-1}$.
 A good understanding of the spectrum of beamstrahlungs photons 
 is needed as they contribute a significant fraction to the 
 production of the charm quark pairs.
 \par
%
\begin{figure}[htb]
\begin{center}
{\includegraphics[width=0.30\linewidth]{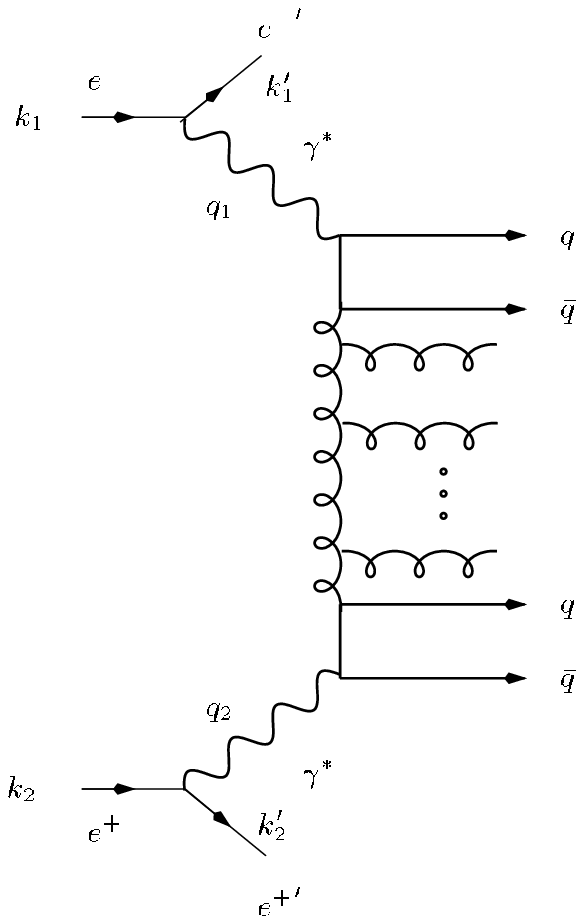}}
{\includegraphics[width=0.40\linewidth]{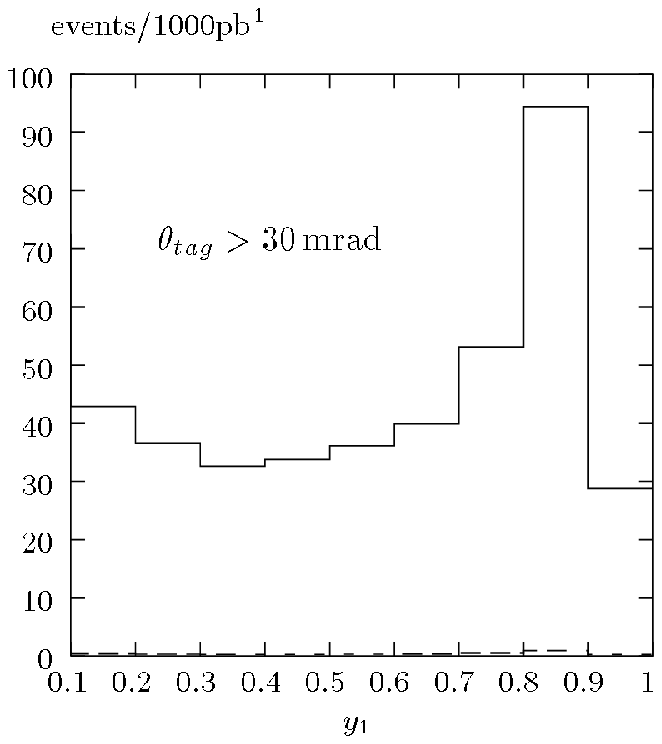}}
\begin{picture}(0,0)
\put(-220,30){\bf(a)}
\put( -40,40){\bf(b)}
\end{picture}
\captive{\label{fig:contr_13} 
 The signature of the BFKL Pomeron in $\sigma_{\gamma^\star\gamma^\star}$,
 from Ref. \protect\cite{BAR-9701}.
 Shown are 
 (a) a sketch of the process
 and
 (b) the expected event rates with (full line), and without 
     (dashed line) the BFKL Pomeron for a specific phase space, see text.
 }
\end{center}
\end{figure}
%
 Since quite some time the search of BFKL signatures at HERA attracted
 broad interest. 
 The observables studied are the behaviour of the proton structure function 
 at small values of $x$ and the production of forward jets. 
 In \epem collisions the BFKL Pomeron would show up in an enhanced 
 total cross section for the scattering of highly virtual photons,
 $\sigma_{\gamma^\star\gamma^\star}$~\cite{BAR-9601,BAR-9701,BRO-9701-BRO-9702}. 
 The diagram of the reaction is shown in Figure~\ref{fig:contr_13}(a).
 In order to freeze the \qsq evolution the two photons are required
 to have similar virtualities. Then the BFKL Pomeron would lead to an 
 enhanced cross section over the two gluon exchange, which is the first 
 significant contribution without BFKL.
 Defining, $y_1=q_2k_1/k_1k_2$, $y_2=q_1k_2/k_1k_2$, 
 $Q_i^2=-q_i^2$, $s=(k_1+k_2)^2$ and 
 $s_0=\sqrt{Q_1^2Q_2^2}/y_1y_2$ the cross section
 expected for the kinematical range, $\ttag>30$ mrad, $\etag>20$ \gev,
 $\log(s/s_0)>2$ and $2.5 < Q_i^2 <200$ \gevsq is about
 $0.3$~pb. This means a yield of ${\cal{O}}(6000)$ events 
 for an integrated luminosity of 20 fb$^{-1}$
 with practically no background from the two gluon exchange 
 process, see Figure~\ref{fig:contr_13}(b).
 The detector requirements are very demanding. In order to observe
 the BFKL signal which dies out like $Q^{-6}$ the instrumentation of the
 mask is a must. In addition the observation of the enhancement relies on
 the ability to detect electrons of relatively low energy of only 20 \gev.
 This is very challenging given the magnitude of the 
 expected machine background discussed above.
 \par
%
\begin{figure}[htb]
\begin{center}
{\includegraphics[width=0.7\linewidth]{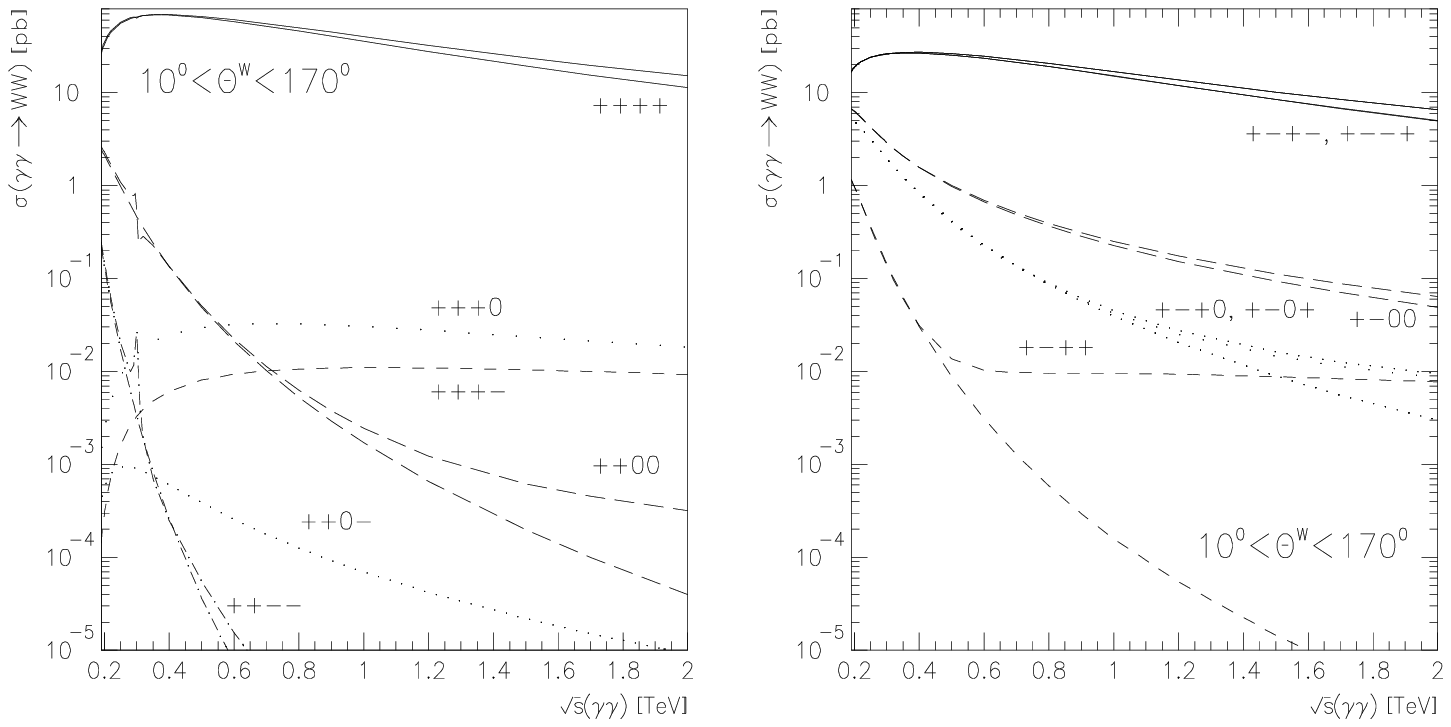}}
\begin{picture}(0,0)
\put(-195,110){\bf(a)}
\put( -40,110){\bf(b)}
\end{picture}
\captive{\label{fig:contr_14} 
 The prospects for $W$ pair production at a future linear 
 collider, from Ref. \protect\cite{JIK-9701}.
 Shown are the Born cross sections and
 the ${\cal{O}}(\alpha)$ cross sections for the reaction
 $\gamma\gamma\rightarrow\WWbar(\gamma)$ for different helicity states
 of the incoming photons and the outgoing $W$ bosons.
 The curves nearest to the helicity symbols denote the 
 ${\cal{O}}(\alpha)$ corrected cross sections.
 }
\end{center}
\end{figure}
%
 Due to the high energy photons produced at a PLC the large cross
 section for $\gamma\gamma\rightarrow\WWbar(\gamma)$ can be 
 exploited.
 Figure~\ref{fig:contr_14} shows the cross sections for $WW$ and
 $WW\gamma$ final states for the Born term and the
 ${\cal{O}}(\alpha)$ corrections in a restricted range in polar angle
 of the $W$ bosons, $10<\theta^W<170$~deg,~\cite{JIK-9701}.
 The cross section at $\sqrt{s_{\GG}}=500$~\gev within this restricted 
 range is $\sigma_{\gamma\gamma}=61$~pb, which
 is to be compared with a cross section of only
 $\sigma_{{\rm ee}}=6.6$~pb in the \epem case at $\sqrt{s_{\ee}}=500$~\gev.
 It is found that the radiative corrections are moderate at 
 $\sqrt{s_{\GG}}=500$~\gev 
 but do strongly depend on $\theta^W$, so special care has to be taken.
 With such a sample of ${\cal{O}}(10^6)$ \WWbar pairs per year
 detailed studies of the anomalous couplings of the $W$ can be performed.
 With the natural order of magnitude of the predicted anomalous couplings
 the standard model cross sections have to be known to better than
 1$\%$ to measure these small numbers.
 Given the high rate and the precise prediction the $W$ pair production is
 in addition a good candidate to monitor the \GG luminosity at a PLC.
 \par
%
\begin{figure}[htb]
\begin{center}
{\includegraphics[width=0.42\linewidth,clip]{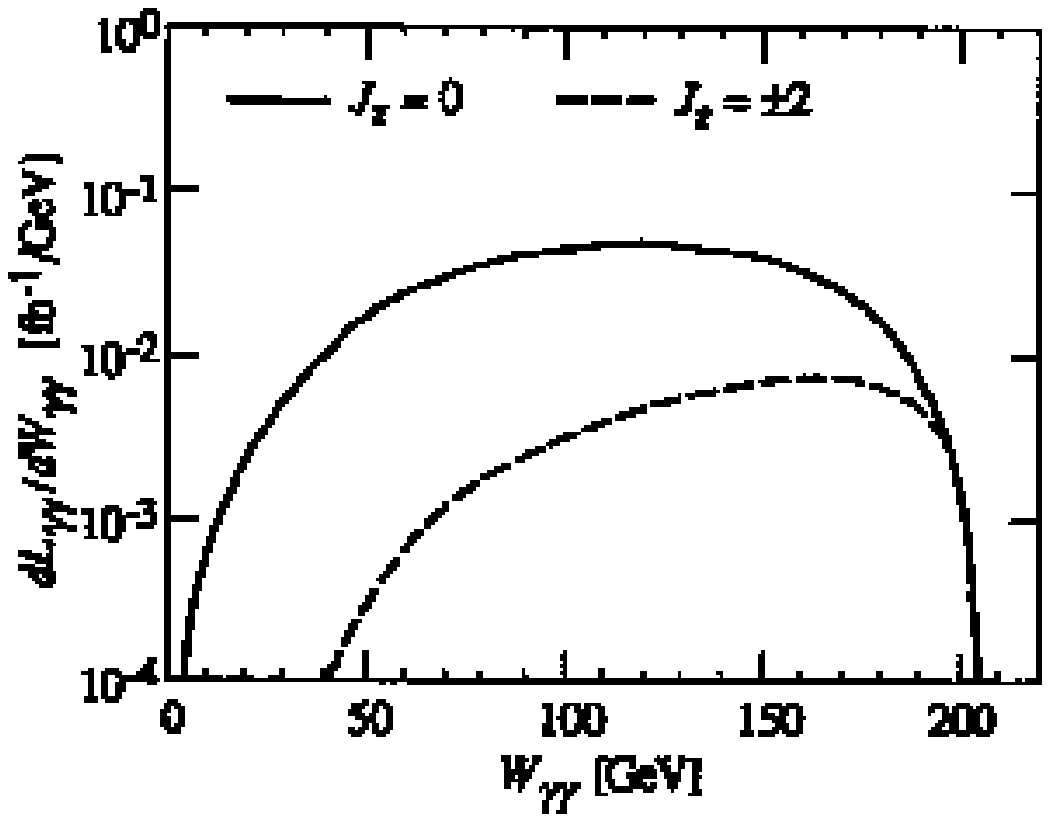}}
{\includegraphics[width=0.42\linewidth,clip]{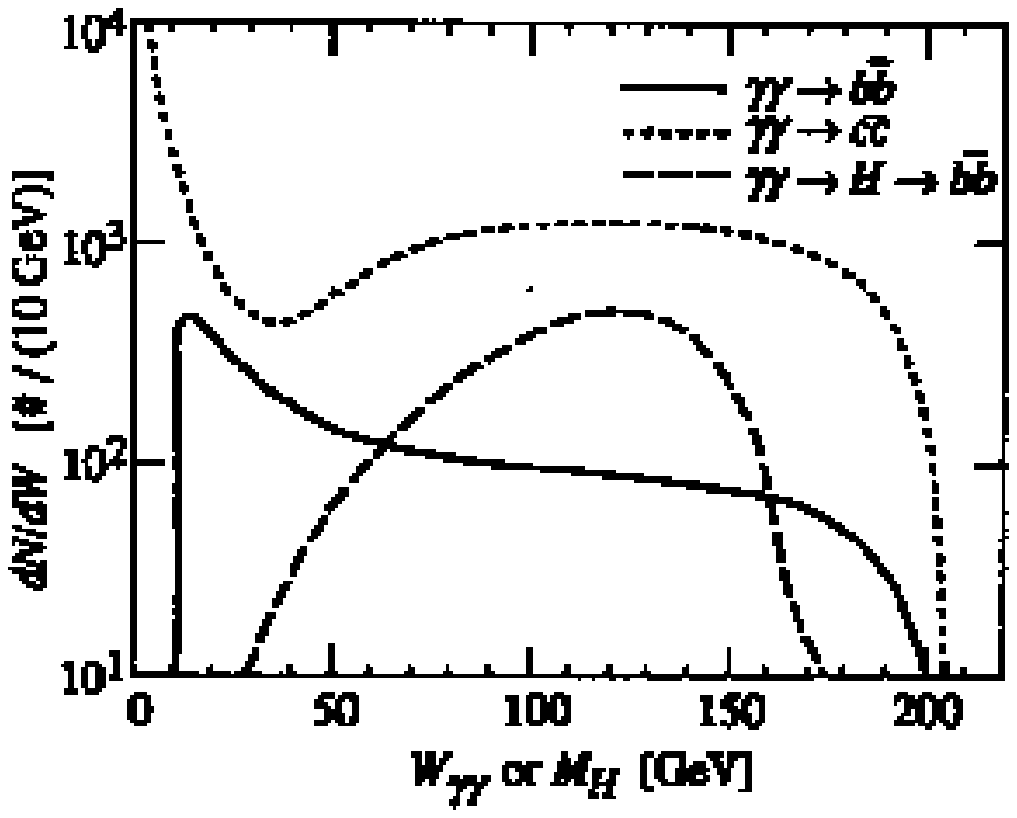}}
{\includegraphics[width=0.42\linewidth,clip]{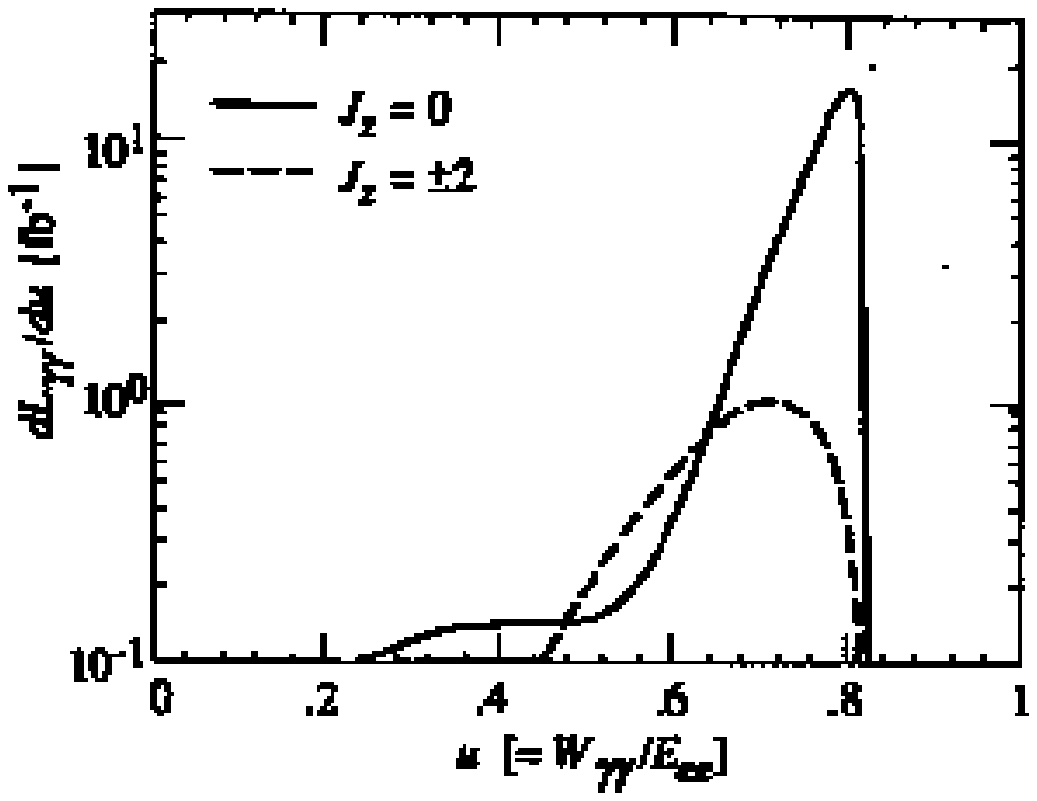}}
{\includegraphics[width=0.42\linewidth,clip]{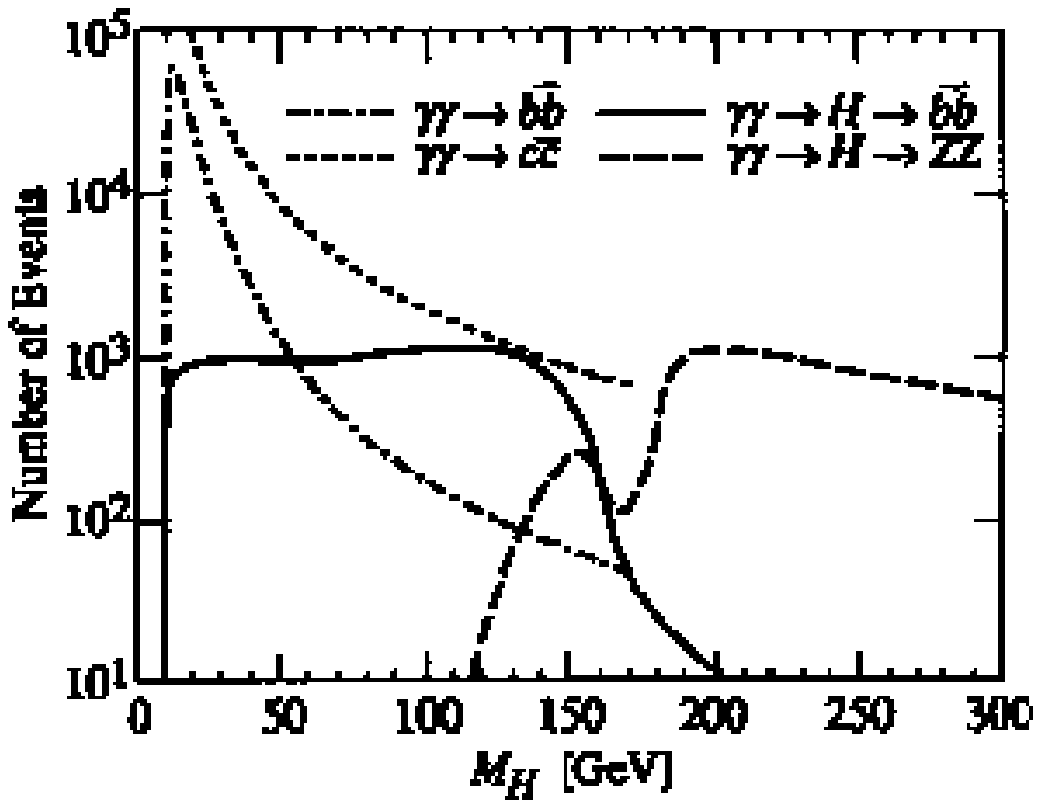}}
\begin{picture}(0,0)
\put(-225,280){\bf(a)}
\put(-110,280){\bf(b)}
\put(-356, 42){\bf(c)}
\put(-150, 42){\bf(d)}
\end{picture}
\captive{\label{fig:contr_15} 
 Higgs production at a future linear collider, from 
 Ref. \protect\cite{BOR-9301}.
 Shown are 
 (a) the photon luminosity spectrum assumed for the Higgs search,
 (b) the expected event rate using the spectrum of (a),
 (c) the photon luminosity spectrum assumed for the measurement of 
     $\Gamma(H\rightarrow\gamma\gamma)$
 and
 (d) the expected event rate  for the measurement of 
     $\Gamma(H\rightarrow\gamma\gamma)$
     using the spectrum of (c).
 }
\end{center}
\end{figure}
%
 Since long the search for the Higgs boson has been performed 
 at various colliders, but without success
 so far. The PLC collider is an ideal place to search for the
 Higgs boson as at such a machine the Higgs boson is produced as an 
 s-channel resonance.
 The most promising channel is $\gamma\gamma\rightarrow H\rightarrow\bbbar$.
 This channel receives background from the non resonant production of 
 \bbbar pairs and also of \ccbar pairs which are misidentified as \bbbar final
 states. To achieve a good signal to noise ratio
 several facts have to be exploited~\cite{BOR-9301}.
 To suppress the continuum production of \bbbar and \ccbar
 a vanishing third component of the total angular
 momentum of the \GG system, $J_z=0$, Figure~\ref{fig:contr_15}(a),
 has to be selected.
 In addition a good tagging efficiency for bottom quarks, 
 $\epsilon_b > 90 \%$, and a good
 suppression of charm quarks of $\epsilon_{c\rightarrow b} <5 \%$ is mandatory.
 Assuming these numbers, together with a mass resolution of 
 $0.1M_H$ (FWHM), a signal with larger than 10 $\sigma$ significance 
 can be established in the range  $80 <M_H <140$ GeV for 
 an integrated luminosity of $10$ fb$^{-1}$~\cite{BOR-9301},
 Figure~\ref{fig:contr_15}(b).
 Once the Higgs has been seen a very fundamental measurement to be performed
 is the determination of the total width, 
 $\Gamma(H\rightarrow\gamma\gamma)$, as it is sensitive to all new particles
 in the loop which couple to the Higgs.
 The results of a feasibility study~\cite{BOR-9301}
 can be seen in Figure~\ref{fig:contr_15}(c,d), where 
 the expected event rates are calculated for a restricted range
 in polar angle for the produced Higgs bosons of $|\cos\theta| <0.7$.
 Assuming a resolution of $\sigma_{M_H} = 0.1 M_H$ the total width
 $\Gamma(H\rightarrow\gamma\gamma)$ can
 be determined with an $\cal{O}$(10\%) error for 
 an integrated luminosity of $10$ fb$^{-1}$~\cite{BOR-9301}.
 \par
%
\begin{figure}[htb]
\begin{center}
{\includegraphics[width=0.40\linewidth,clip]{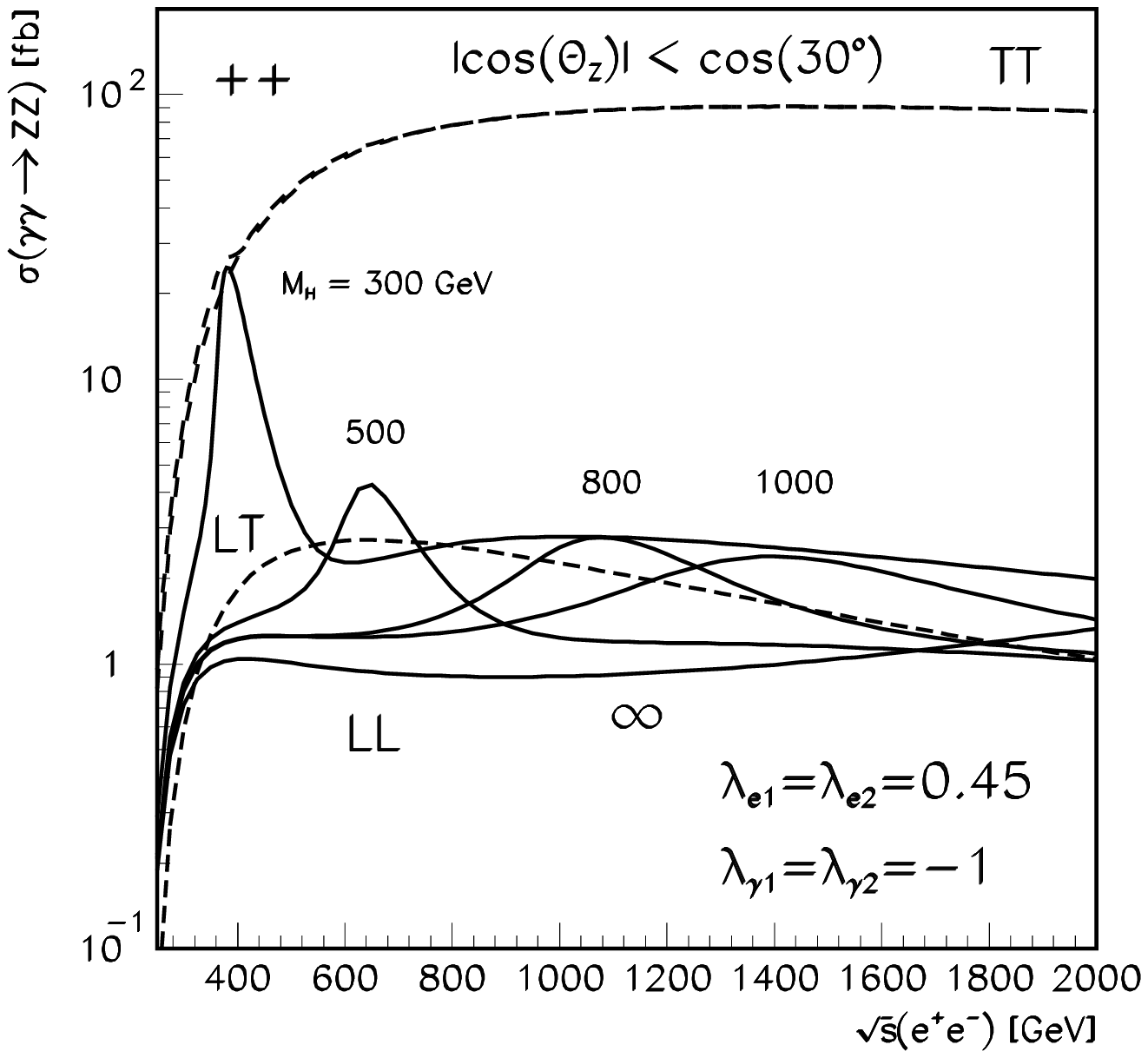}}
{\includegraphics[width=0.40\linewidth,clip]{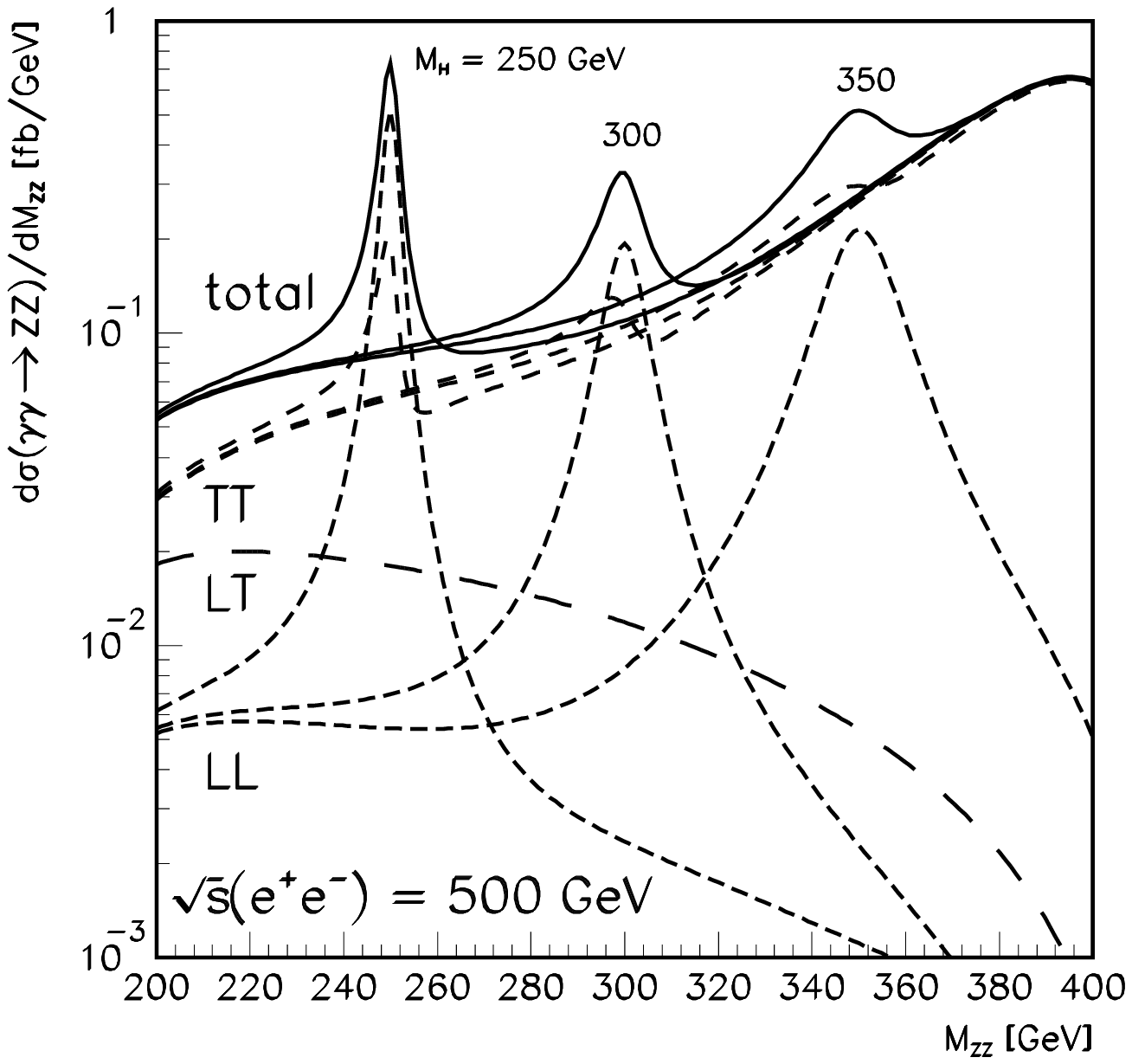}}
\begin{picture}(0,0)
\put(-225,135){\bf(a)}
\put( -35,135){\bf(b)}
\end{picture}
\captive{\label{fig:contr_16} 
 Prospects for the Higgs search in the rection
 $\gamma\gamma\rightarrow H\rightarrow Z Z$,
 from Ref. \protect\cite{JIK-9702}.
 Shown are 
 (a) the cross sections as a function of the \epem centre-of-mass energy and  
 (b) the invariant mass distributions at $\sqrt{s_{\rm ee}} = 500$~\gev
  for $Z$ pair production in \GG collisions
  at a future linear collider using the photon spectrum of a PLC.
 The curves are for different helicities of the $Z$ bosons and different
 masses of a hypothetical Higgs boson.
 }
\end{center}
\end{figure}
%
 Another interesting channel is the reaction
 $\gamma\gamma\rightarrow H\rightarrow Z Z$. 
 As can be seen from Figure~\ref{fig:contr_16}
 the cross section strongly depends on the helicities of the $Z$ bosons.
 In this channel Higgs signals up to $M_H=350$ GeV can be observed. For 
 higher masses the background from the continuum $Z_T Z_T$ production 
 is too high~\cite{JIK-9702}.
%
%
\section{Summary}
\label{sec:concl}
 The linear collider is an unique place to investigate two photon
 physics at the highest energies. 
 Due to the high centre-of-mass energy of the photon-photon system
 especially in the case of the \GG collider new 
 channels like Higgs bosons, $W$ pairs and $Z$ pairs are
 open to be copiously produced in reactions of two photons.
 This opens a very rich field of interesting measurements to be performed.
 The tagging of electrons down to the lowest possible 
 angles is a challenging task, but it is mandatory in order
 to achieve overlap in \qsq with the measurements of the photon
 structure function \ft at LEP.
 In this report the available information on the prospects for the 
 measurement of the structure function \ft at a future linear collider
 was extended by a detailed discussion of the prospects 
 for the measurement of the \qsq evolution of \ft.
 In all physics channels a careful determination of the
 \GG, \eG and \ee luminosity distributions is essential.
 Lots of work is in front of us to bring a linear collider to life, but 
 it should be fun and the physics potential is certainly worth the effort.
%
%
\par
\section*{Acknowledgements}
 I wish to thank Albert De Roeck for useful discussions and 
 Bernd Surrow for presenting the results on my behalf at the conference.
 \par
%
%

%
%
\end{document}